\documentclass[12pt]{mn2e}
\usepackage{graphicx}
\usepackage{epsf}
\usepackage{lscape}
\usepackage{longtable}
\usepackage{rotating}
\usepackage{color}
\pdfoutput=1

\newcommand{\aap}{A\&A}

\newcommand{\apj}{ApJ}
\newcommand{\apjl}{ApJ}

\newcommand{\mnras}{MNRAS}

\definecolor{grey}{rgb}{0.7,0.7,0.7}

\begin{document}
\topmargin -0.5in 

\title[The effect of nebular emission]{Theoretical predictions for the effect of nebular emission on the broad band photometry of high-redshift galaxies}

\author[Stephen M. Wilkins et al.]  
{
Stephen M. Wilkins$^{1,2}$\thanks{E-mail: s.wilkins@sussex.ox.ac.uk}, William Coulton$^{2}$, Joseph Caruana$^{2,3}$, Rupert Croft$^{2,4}$, \newauthor Tiziana Di Matteo$^{2,4}$, Nishikanta Khandai$^{4,5}$, Yu Feng$^{4}$, Andrew Bunker$^{2}$, Holly Elbert$^{2}$  \\
$^1$\,Astronomy Centre, Department of Physics and Astronomy, University of Sussex, Brighton, BN1 9QH, U.K. \\
$^2$\,Department of Physics, Denys Wilkinson Building, University of Oxford, Keble Road, OX1 3RH, U.K. \\
$^3$\,Leibniz Institute for Astrophysics, An der Sternwarte 16, 14482 Potsdam, Germany\\
$^4$\,McWilliams Center for Cosmology, Carnegie Mellon University, 5000 Forbes Avenue, Pittsburgh, PA 15213, U.S.A.\\
$^5$\,Brookhaven National Laboratory, Department of Physics, Upton, NY 11973, U.S.A.
}
\maketitle

\begin{abstract}
By combining optical and near-IR observations from the {\em Hubble Space Telescope} with near-IR photometry from the {\em Spitzer Space Telescope} it is possible to measure the rest-frame UV-optical colours of galaxies at $z=4-8$. The UV - optical spectral energy distribution of star formation dominated galaxies is the result of several different factors. These include the joint distribution of stellar masses, ages, and metallicities (solely responsible for the pure stellar spectral energy distribution), and the subsequent reprocessing by dust and gas in the interstellar medium. Using a large cosmological hydrodynamical simulation ({\em MassiveBlack-II}) we investigate the predicted spectral energy distributions of galaxies at high-redshift with a particular emphasis on assessing the potential contribution of nebular emission. We find that the average (median) pure stellar UV-optical colour correlates with both luminosity and redshift such that galaxies at lower-redshift and higher-luminosity are typically redder. Assuming the escape fraction of ionising photons is close to zero, the effect of nebular emission is to redden the UV-optical $1500-V_{w}$ colour by, on average, $0.4\,{\rm mag}$ at $z=8$ declining to $0.25\,{\rm mag}$ at $z=4$. Young and low-metallicity stellar populations, which typically have bluer pure stellar UV-optical colours, produce larger ionising luminosities and are thus more strongly affected by the reddening effects of nebular emission. This causes the distribution of $1500-V_{w}$ colours to narrow and the trends with luminosity and redshift to weaken. The strong effect of nebular emission leaves observed-frame colours critically sensitive to the redshift of the source. For example, increasing the redshift by $0.1$ can result in observed frame colours changing by up to $\sim 0.6$. These predictions reinforce the need to include nebular emission when modelling the spectral energy distributions of galaxies at high-redshift and also highlight the difficultly in interpreting the observed colours of individual galaxies without precise redshift information. 
\end{abstract} 


\section{Introduction}

The availability of deep {\em Hubble Space Telescope} surveys utilising the Advanced Camera for Surveys (ACS) and more recently Wide Field Camera 3 (WFC3) means it is now possible to routinely identify galaxies at very-high redshift, with large ($>50$) samples identified to $z\approx 8$ (e.g. Oesch et al. 2010a, Bouwens et al. 2010a, Bunker et al. 2010, Wilkins et al. 2010, Finkelstein et al. 2010, Wilkins et al. 2011a, Lorezoni et al. 2011, Bouwens et al. 2011, Lorenzoni et al. 2013, McLure et al. 2013, Schenker et al. 2013) and a few candidates now identified at $z>10$ (e.g. Oesch et al. 2012a, Bouwens et al 2012b, Coe et al. 2012, Oesch et al. 2013, Ellis et al. 2013). 

While {\em Hubble} ACS and WFC3 observations (which probe the rest-frame UV continuum at $z>3$) alone allow us to learn a great deal about high-redshift galaxies, including the UV luminosity function (e.g. Oesch et al. 2010a, Bouwens et al. 2010a, Wilkins et al. 2011a, Lorezoni et al. 2011, Bouwens et al. 2011, Lorenzoni et al. 2013, McLure et al. 2013, Schenker et al. 2013), the UV continuum slope (e.g. Stanway et al. 2005, Bunker et al. 2010, Bouwens et al. 2010b, Wilkins et al. 2011b, Wilkins et al. 2013a, Bouwens et al. 2013), and UV morphologies (e.g. Oesch et al. 2010b), by combining them with {\em Spitzer} Infrared Array Camera (IRAC) photometry it is possible to also probe the rest-frame optical emission. This is extremely difficult (given the lower sensitivity of the IRAC observations) for all but the brightest individual objects. However, by stacking large samples of galaxies together it becomes possible to robustly probe the average SEDs of even the faintest galaxies (e.g. Eyles et al. 2005, Labb{\'e} et al. 2010, Gonz{\'a}lez et al. 2011, Labb{\'e} et al. 2012). 

The UV/optical spectral energy distribution of a star formation dominated galaxy are affected by a complex mixture of different factors including the joint distribution of stellar masses, ages, and metallicities, dust, and nebular emission, many of which are closely coupled. The large number of effects makes it difficult to {\em ab initio} interpret observations, especially at high-redshift, where typically only broadband photometry is available, in the context of any of these individual quantities.   

In this paper we use a state-of-the-art cosmological hydrodynamical simulation of structure formation ({\em MassiveBlack}-II) to investigate the UV-optical colours of high-redshift galaxies and in particular the effect of nebular emission thereupon. This paper is organised as follows: in Section \ref{sec:factors} we discuss in turn the various factors affecting the rest-frame UV/optical colours of high-redshift galaxies. In Section \ref{sec:sims} we present predictions from our large cosmological hydrodynamic simulation {\em MassiveBlack}-II. In Section \ref{sec:stellar_masses} we describe how strong nebular emission makes robust estimates of galaxy stellar masses difficult. Finally, in Section \ref{sec:c} we present our conclusions. Magnitudes are calculated using the AB system (Oke \& Gunn 1983). Throughout this work we assume a Salpeter (1955) stellar initial mass function (IMF), i.e.: $\xi(m)={\rm d}N/{\rm d}m\propto m^{-2.35}$. 

\subsection{Filters used to probe the UV-optical SEDs}

\begin{figure*}
\centering
\includegraphics[width=40pc]{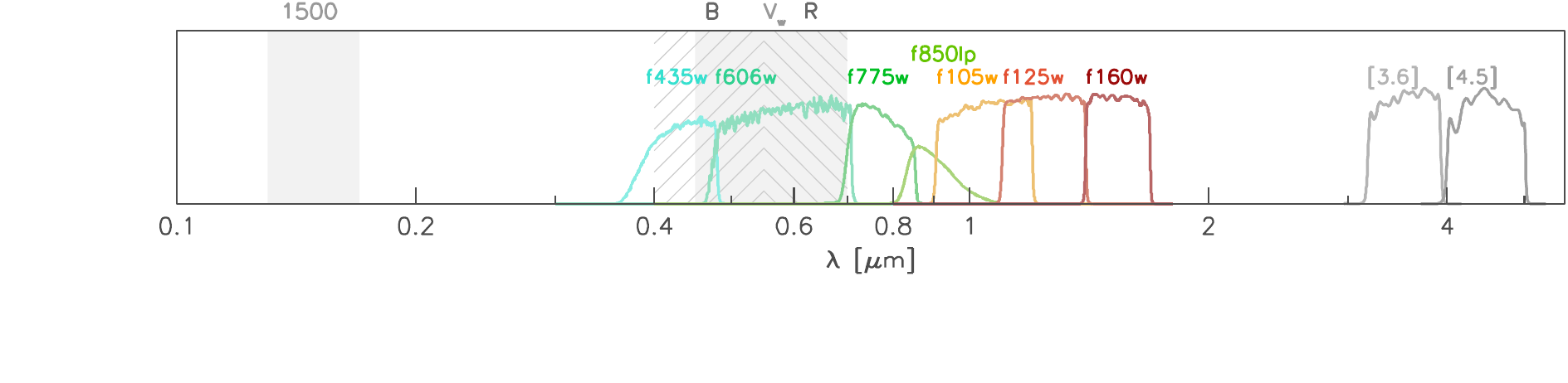}
\caption{The transmission profiles of the various filters used throughout this study.}
\label{fig:filters}
\end{figure*}

\begin{figure*}
\centering
\includegraphics[width=20pc]{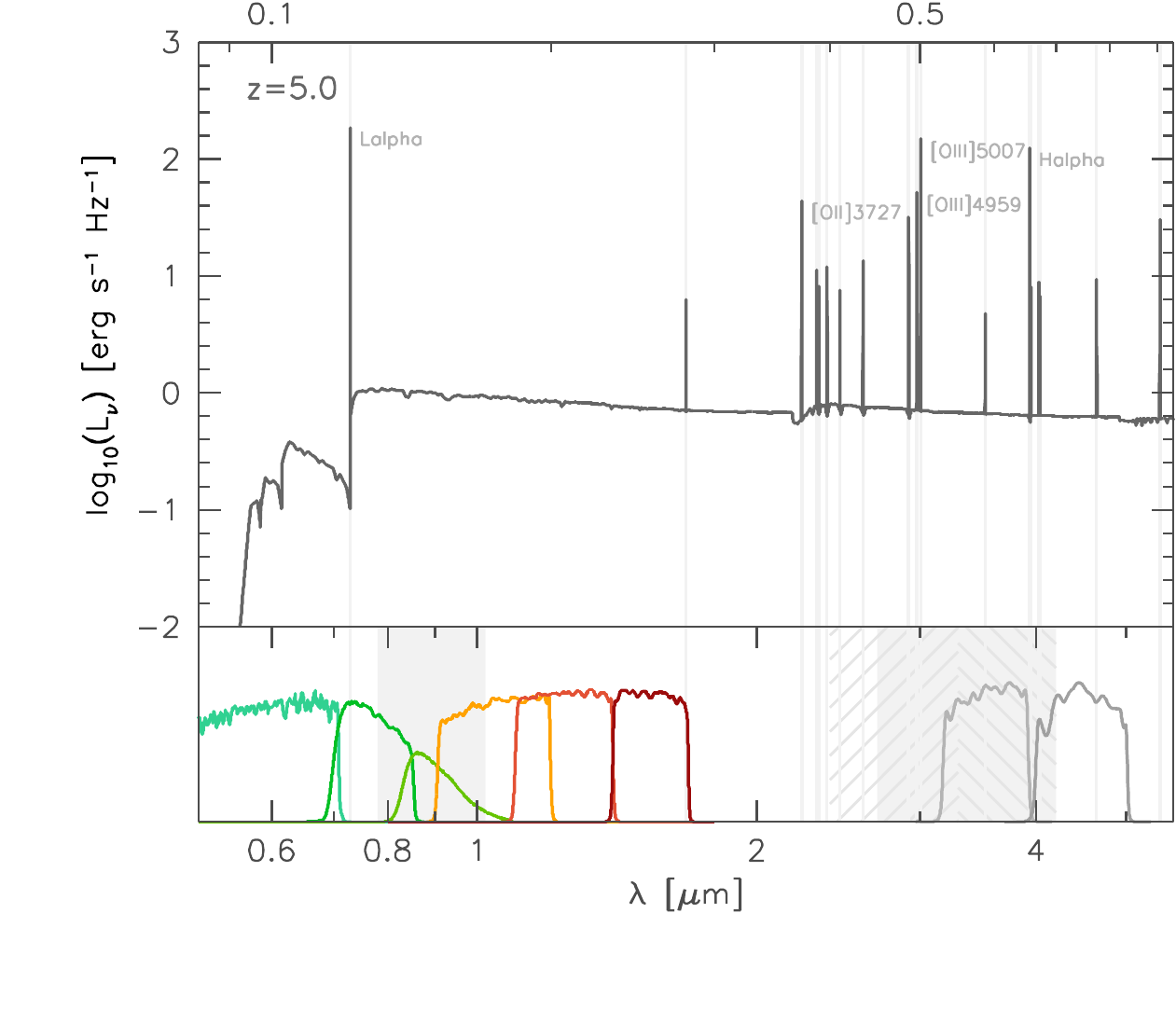}
\includegraphics[width=20pc]{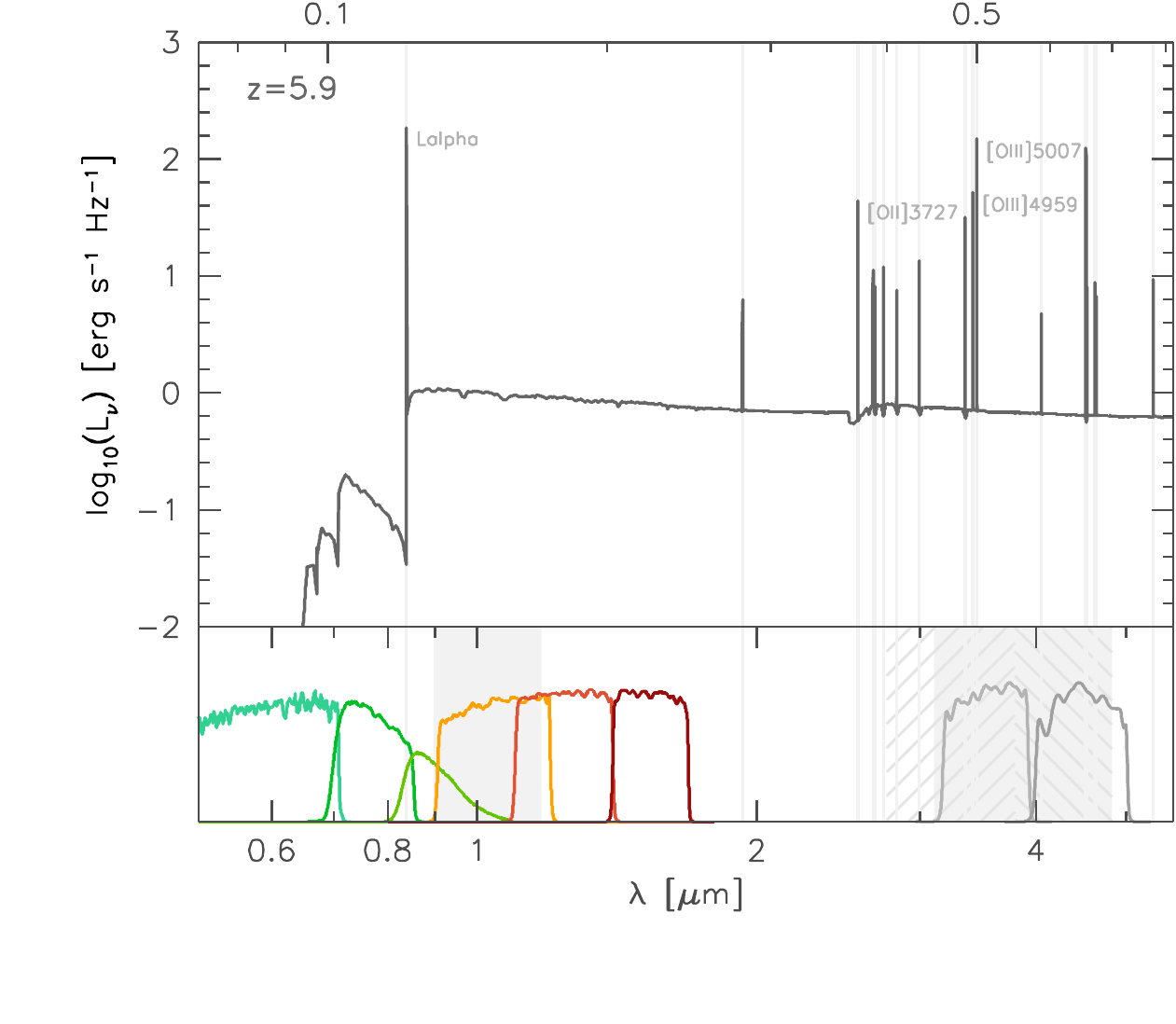}
\includegraphics[width=20pc]{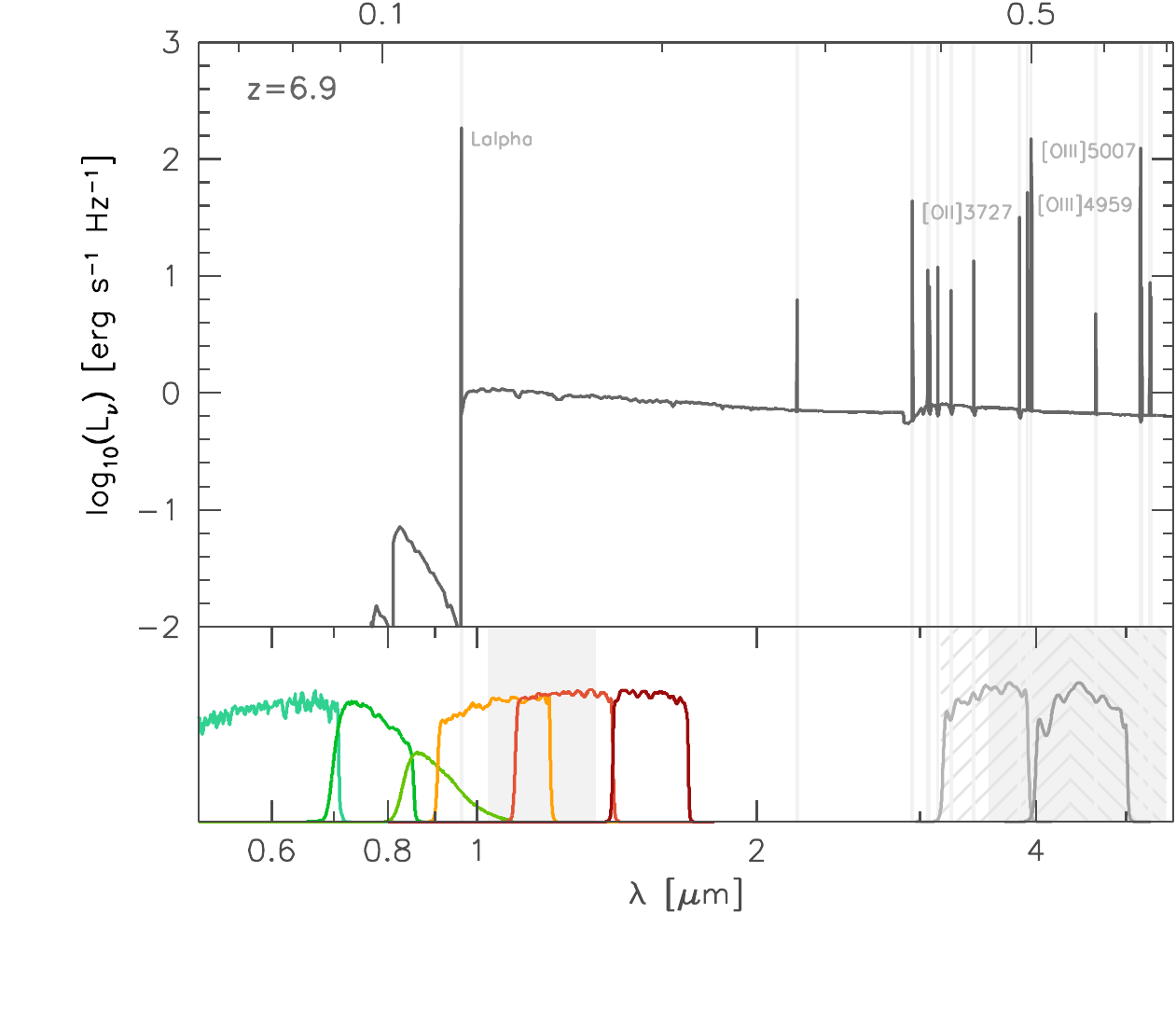}
\includegraphics[width=20pc]{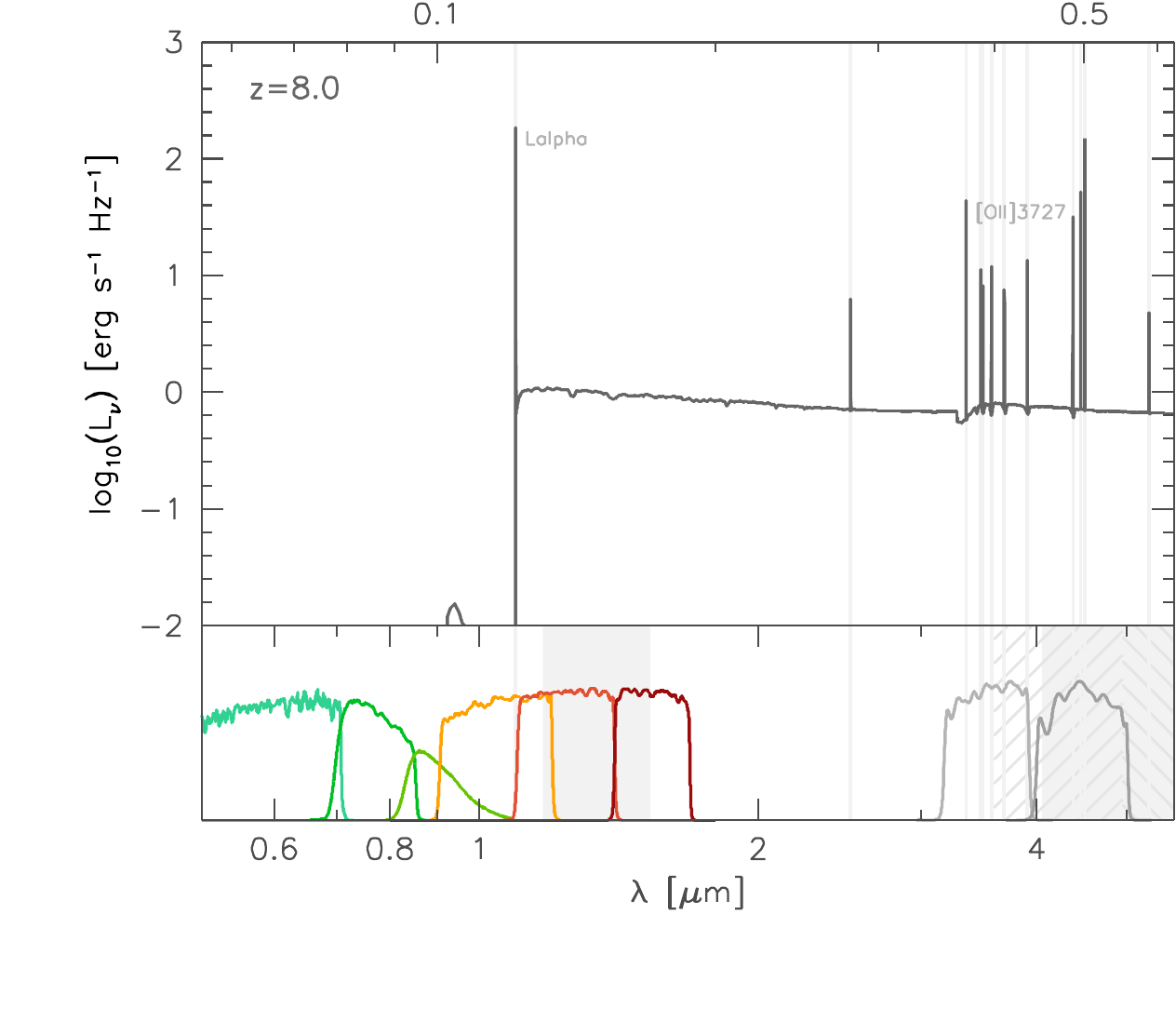}
\caption{The spectral energy distribution of a stellar population which has formed stars at a constant rate for $100\,{\rm Myr}$ with $z=0.004$ at $z\in\{5.0,5.9,6.9,8.0\}$. For clarity only emission lines with $f_{x}>0.2\times f_{{\rm H}\beta}$ are shown (and only those with $f_{x}>f_{{\rm H}\beta}$ are labelled). The lower panel of each figure shows various {\em Hubble}/ACS (V$_{f606w}$, $i_{f775w}$, and $z_{f850lp}$), {\em Hubble}/WFC3 (Y$_{f105w}$, J$_{f125w}$, and H$_{f160ww}$), and {\em Spitzer}/IRAC ([3.6] and [4.5]) transmission curves. The shaded regions denote two artificial rest-frame bandpasses ($1500$ and $V_{w}$) and the two hatched regions denote the $B$ and $R$ bandpasses.}
\label{fig:SEDs}
\end{figure*}

Throughout this work we make use of several {\em Hubble} and {\em Spitzer} filters, including: {\em Hubble}/ACS ($B_{f435w}$, $V_{f606w}$, $i_{f775w}$, and $z_{f850lp}$), {\em Hubble}/WFC3 ($Y_{f105w}$, $J_{f125w}$, and $H_{f160w}$), and {\em Spitzer}/IRAC ([3.6] and [4.5]) filters. We also introduce four rest-frame bandpasses ($1500$\footnote{Defined as $T_{\lambda}=[0.13<\lambda/\mu m<0.17]^{5}$.}, $B$\footnote{Defined as $T_{\lambda}=[0.40<\lambda/\mu m<0.55]^{5}.$}, $V_{w}$\footnote{Defined as $T_{\lambda}=[0.45<\lambda/\mu m<0.70]^{5}$.}, and $R$\footnote{Defined as $T_{\lambda}=[0.55<\lambda/\mu m<0.70]^{5}.$})\footnote{Here we employ the Iverson bracket notation to define the bandpass, $[A]=1.0$ when A is {\em True} and $0.0$ otherwise.}. These rest-frame bandpasses allow us to consistently compare the properties of galaxies at different redshifts. The simple shape of these filters is chosen for convenience and the transmission profiles of all these filters are shown in Figure \ref{fig:filters}. Figure \ref{fig:SEDs} shows the observed frame SED of a star forming galaxy at $z\in\{5.0,5.9,6.9,8.0\}$ in relation to this filter set.

\section{Factors affecting the UV-optical colours of star forming galaxies}\label{sec:factors}

The observed spectral energy distributions (SEDs) of galaxies are formed from the intrinsic stellar and AGN SEDs with reprocessing by dust and gas (both in the local ISM and IGM). The intrinsic SED of a stellar population (i.e. the {\em pure} stellar SED) is determined by the {\em joint} distribution of stellar masses, ages, and metallicities. To demonstrate the effect of various changes to the star formation and metal enrichment histories we utilise the {\sc Pegase.2} (Fioc \& Rocca-Volmerange 1997,1999) stellar population synthesis (SPS) model. We first, in \S\ref{sec:factors.ma}-\ref{sec:factors.Z}, describe how the rest-frame intrinsic pure stellar UV-optical $1500-V_{w}$ and optical $B-R$ colours are affected by the properties of the stellar population (distribution of masses, ages, and metallicities). In \S\ref{sec:factors.dust} we extend this to include the effect of dust and critically in \S\ref{sec:factors.neb} we discuss the effect of nebular emission. 

It is also important to stress that the predicted SED, for a given IMF, star formation history, and metal enrichment history, is also sensitive to the choice of SPS model. In Appendix \ref{app:SPS} we investigate how changing the SPS model affects the predicted UV-optical colours.

\subsection{Distribution of stellar mass and ages}\label{sec:factors.ma}

The SEDs of individual stars vary strongly with both stellar mass and evolutionary stage (and therefore age). As such the intrinsic SED of a composite stellar population is predominantly determined by the joint distribution of stellar masses and ages. This, in turn, is determined by both the initial mass function (IMF) and the star formation history (SFH).

\begin{figure}
\centering
\includegraphics[width=20pc]{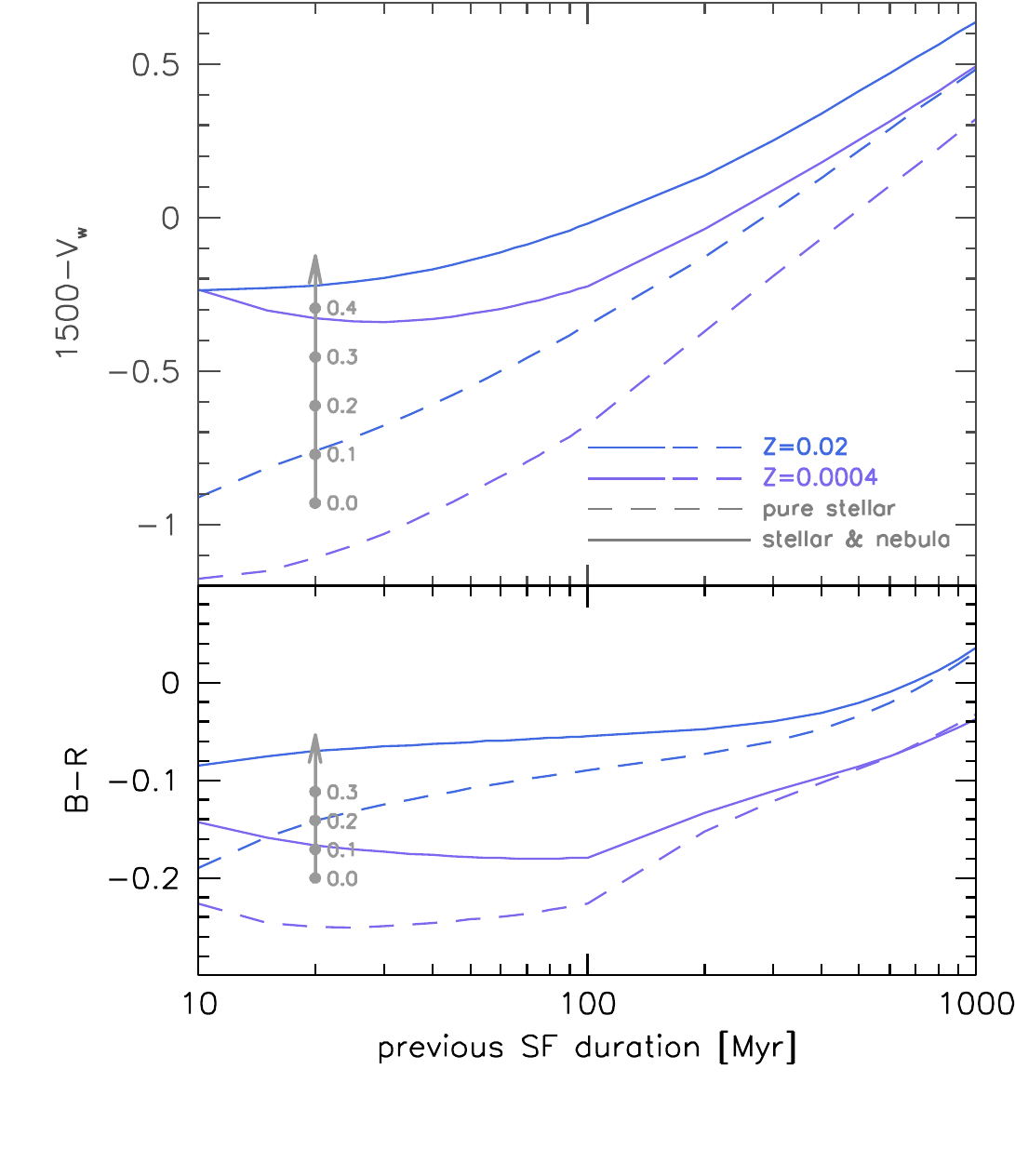}
\caption{The predicted (using the {\sc Pegase.2} SPS code) UV-optical ($1500-V_{w}$) and optical $B-R$ (lower panel, note the difference in scale) colours as a function of the previous duration of (constant) star formation. Colours are shown for both two metallicities ($Z\in\{0.02, 0.0004\}$) and both for a pure stellar SED (dashed lines) and the SED including a nebular contribution (solid lines). The two arrows show the effect of $A_{V}=0.5$ dust attenuation (with labels denoting intermediate values of $A_{V}$) assuming the Calzetti et al. (2000) reddening law.}
\label{fig:combined_rest}
\end{figure}

\subsubsection{Star formation history}\label{sec:factors.ma.sfh}

In general, stellar populations with protracted star formation histories will contain a higher proportion of low-mass stars (as many of the original high-mass stars will have evolved off the main-sequence) causing the SED of the population to redden relative to a younger population. The sensitivity of the UV/optical colours to the recent star formation history can be seen in Figure \ref{fig:combined_rest}; this shows the sensitivity of both the rest-frame $1500-V_{w}$ (UV-optical) and $B-R$ (optical) colours to the duration of previous (constant) star formation.  Increasing the duration of previous star formation from $10\to 1000\,{\rm Myr}$ causes the pure stellar $1500-V_{w}$ colour to redden by $\sim 1.4\,{\rm mag}$. The effect on the optical $B-R$ colour is more subtle, with the colour increasing by only $\sim 0.2\,{\rm mag}$ as the previous duration of star formation is increased from $10\to 1000\,{\rm Myr}$.

Assuming an increasing SFH, which is likely to be more representative of high-redshift star forming galaxies (see for example the predictions of Finlator et al. 2011), will suppress the evolution of the $1500-V_{w}$ colour as the SED is remains dominated by the most massive stars. In contrast if we instead consider the colour evolution of an instantaneous burst over the same time period ($10\to 1000\,{\rm Myr}$) the $1500-V_{w}$ and $B-R$ colours redden by $\sim 3.5\,{\rm mag}$ and $\sim 0.5\,{\rm mag}$ respectively.

\subsubsection{Initial mass function}\label{sec:factors.ma.imf}

Changing the choice of initial mass function will also effect the distribution of stellar masses. As such it can potentially have a significant affect on the colours of a stellar population. Changes to the low-mass ($<0.5\,{\rm M_{\odot}}$) end of the IMF, have only a small effect on the shape of the SED as these stars contribute only a small fraction of the total luminosity (especially in the UV/optical) of actively star forming galaxies. On the other hand, changes to the high-mass end, will affect the mass-distribution of luminous massive stars. 

\begin{figure}
\centering
\includegraphics[width=20pc]{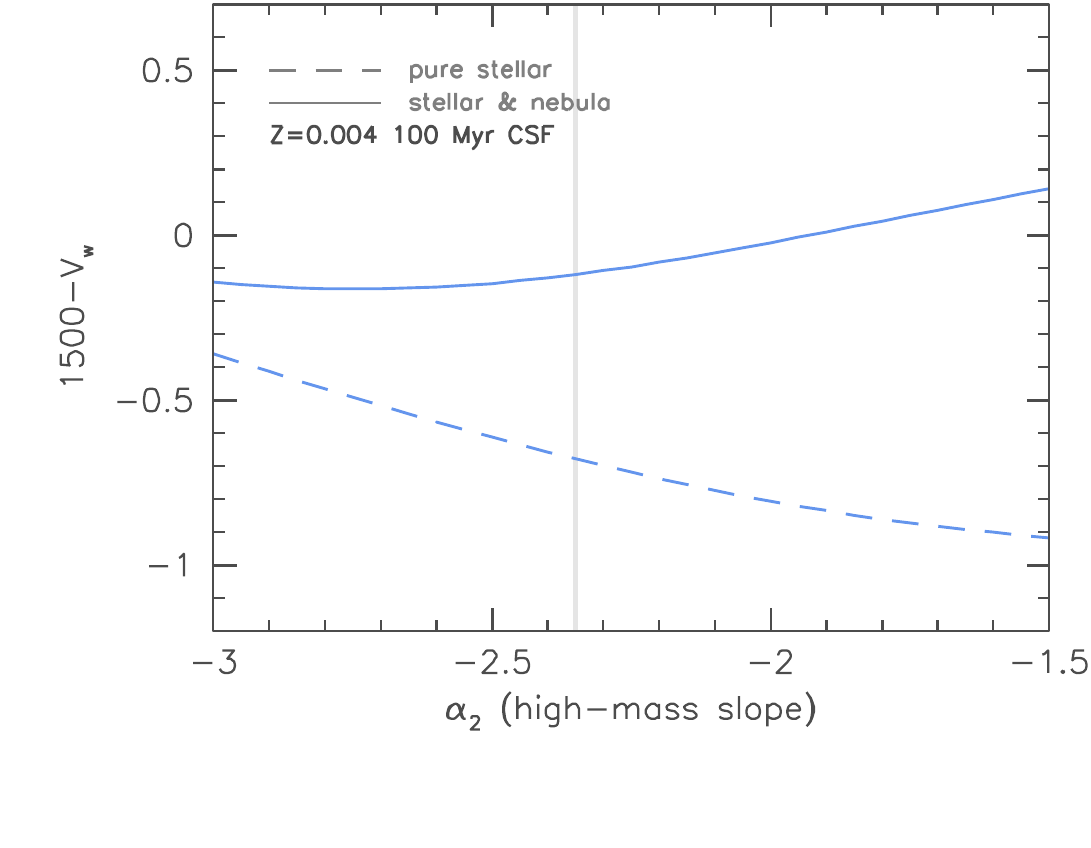}
\caption{The predicted UV-optical ($1500-V_{w}$) colour as a function of the choice of IMF high-mass slope ($\alpha_{2}$) for both a pure stellar SED (dashed lines) and the SED including a nebular contribution (solid lines). In each case a metallicity of $Z=0.004$ and a $100\,{\rm Myr}$ duration of previous star formation are assumed. The vertical line denotes the Salpeter (1955) IMF ($\alpha_{2}=-2.35$).}
\label{fig:combined_rest_IMF}
\end{figure}

The high-mass IMF can be most simply parameterised as a power law, i.e. $\xi(m>0.5\,{\rm M_{\odot}})={\rm d}N/{\rm d}m\propto m^{\alpha_{2}}$, where $\alpha_{2}$ is the high-mass slope (which for the Salpeter 1955 IMF would be $\alpha_{2}=-2.35$). Increasing $\alpha_{2}$ increases the relative proportion of very-high mass stars resulting in a bluer $1500-V_{w}$ colour, as can be seen in Figure \ref{fig:combined_rest_IMF}. Changing $\alpha_{2}$ from $-2.35$ to $-1.5$ results in the pure stellar colour decreasing by $\sim 0.2\,{\rm mag}$. 

\subsection{Metal enrichment history}\label{sec:factors.Z}

Stars of similar mass and age with lower metallicities generally have higher effective-temperatures and thus bluer UV/optical colours (see also Wilkins et al. 2012, Wilkins et al. 2013a). A stellar population with a similar star formation history, but lower metallicity, will also then tend to have bluer UV-optical colours. This can be seen in Figure \ref{fig:combined_rest}, where the UV/optical colours ($1500-V_{w}$, $B-R$) are shown for two metallicities ($Z\in\{0.02, 0.0004\}$). While (assuming the same duration of previous star formation) the high metallicity population is always redder, the difference is sensitive to the previous star formation duration (and thus the distribution of stellar masses and ages).

\subsection{Dust}\label{sec:factors.dust}

Throughout the UV to NIR dust acts to preferentially absorb light at shorter wavelengths. One effect of dust is then to cause the observed colour to redden relative to the intrinsic colour, i.e. $(m_{a}-m_{b})^{\rm obs} = (m_{a}-m_{b})^{\rm int}+E(B-V)(k_{a}-k_{b})$ where $\lambda_{a}<\lambda_{b}$, $E(B-V)\ge 0.0$ and $k_{a}>k_{b}$. The extent of the reddening due to dust is then a product of the attenuation/reddening curve $k(\lambda)$ and a measure of the total attenuation (often expressed by the colour excess $E(B-V)$). 

Assuming an SMC-like curve (Pei et al. 1992), which is favoured at high-redshift by recent observations (e.g. Oesch et al. 2012b), an optical attenuation of $A_{V}=0.4\,{\rm mag}$ (which corresponds to $A_{1500}\approx 1.76\,{\rm mag}$) will redden the $1500-V_{w}$ and $B-R$ colours by $\approx 1.36$ and $\approx 0.15$ respectively (assuming no distinction in the effect of dust between nebular and stellar emission).

If instead we assume a Calzetti et al. (2000) starburst curve (and again assume no distinction in the effect of dust between nebular and stellar emission) the same $V$-band attenuation (i.e. $A_{V}=0.4\,{\rm mag}$, which corresponds to $A_{1500}\approx 1.03\,{\rm mag}$) reddens the $1500-V_{w}$ colour by $\approx 0.63\,{\rm mag}$ and the $B-R$ colour by only $\approx 0.12\,{\rm mag}$. However, Calzetti et al. (2000) found that the nebular emission of intensely star forming galaxies is more strongly affected by dust than the stellar emission (at the same wavelength) with a the relationship between the colour excess of the nebular and stellar emission of $E(B-V)_{\rm stellar}=(0.44\pm 0.03)\times E(B-V)_{\rm nebular}$. A significant consequence of assuming the Calzetti et al. (2000) is that there is no longer a unique mapping between the intrinsic and observed colours for a given attenuation.

While in this work we are more concerned with effect of nebular emission on the intrinsic photometry it is however worth noting that the similar consequences of dust attenuation, the star formation history and metallicity make interpreting the observed broad-band colours of stellar populations in the context of these quantities extremely challenging.


\subsection{Nebular emission}\label{sec:factors.neb}

Ionising radiation, which is produced predominantly by hot, young ($<10\,{\rm Myr}$), massive stars ($>30\,{\rm M_{\odot}}$)\footnote{This also makes the presence of nebular line emission a powerful diagnostic of unobscured ongoing star formation.}, is potentially reprocessed by gas in the ISM into nebular (line and continuum) emission. At high-redshift ($z>4$), galaxy formation models (see Section \ref{sec:sims}) suggest that virtually all galaxies continue to actively form, or have recently formed, stars\footnote{Even if this were not the case it is likely that many of the galaxies currently {\em observed} (at these redshifts) would be actively forming stars by virtue of being rest-frame UV selected.}. Assuming the the escape fraction of ionising photons ($X_{f}$) is small these galaxies are likely to contain strong nebular line emission.

To include the effect of nebular emission on our predictions we use the number of ionising photons predicted (by {\sc Pegase.2}) to determine the fluxes in the Hydrogen recombination lines. Fluxes in {\em non}-Hydrogen lines are determined using the metallicity dependent conversions of Anders \& Fritze et al. (2003)\footnote{{\sc Pegase.2} can, without modification, output emission line fluxes however it does not take account variations in the line ratios due to metallicity.}. Throughout this analysis we assume that the escape fraction ($X_{f}$) is zero. This assumption allows us, when combined with the pure stellar colours, to explore the full range of potential colours.

\subsubsection{Effect on rest-frame UV-optical colours}\label{sec:factors.neb.rest}

Due to the number of strong emission lines (e.g. O[II], $H_{\beta}$, O[III], and $H_{\alpha}$) in the rest-frame optical, unless the bandpass used to probe the UV encompasses Lyman-$\alpha$, the effect of nebular emission will be to redden the UV-optical colour relative to that of the pure stellar colour. This can be seen in Figure \ref{fig:combined_rest} where both the pure stellar and the stellar with nebular emission colours are shown (as function of previous duration of star formation). 

For a $100\,{\rm Myr}$ duration of previous constant star formation and $Z=0.02$, the effect of including nebular emission is to redden the $1500-V_{w}$ colour by $\sim 0.3$ (see Figure \ref{fig:combined_rest}). However, the relative effect of nebular emission changes with both metallicity and previous star formation duration. For protracted ($>1000\,{\rm Myr}$) constant star formation, the effect decreases to $\sim 0.1\,{\rm mag}$ (if there has been no star formation for $>10\,{\rm Myr}$ the contribution of nebular emission will also fall to virtually zero). This variation is predominantly due to the sensitivity to the ratio of ionising photon flux to the optical flux which is itself sensitive to the (joint) distribution of stellar masses, ages, and metallicities. The variation with metallicity is in part also due to the metallicity dependent line ratios.

The inclusion of nebular emission can (for actively star forming populations) also dramatically modify the trend with the high-mass slope of the IMF compared to the pure stellar case (\S\ref{sec:factors.ma.imf}), as can be seen in Figure \ref{fig:combined_rest_IMF}. When nebular emission is included the trend between the $1500-V_{w}$ colour and high-mass slope reverses: stellar populations with shallower IMFs (high-mass biased) have redder $1500-V_{w}$ colours. This is again a result of the increased proportion of massive hot stars which produce large amounts of ionising radiation.

\subsubsection{Redshift sensitivity and effect on observed colours}\label{sec:factors.neb.obs}

\begin{figure}
\centering
\includegraphics[width=20pc]{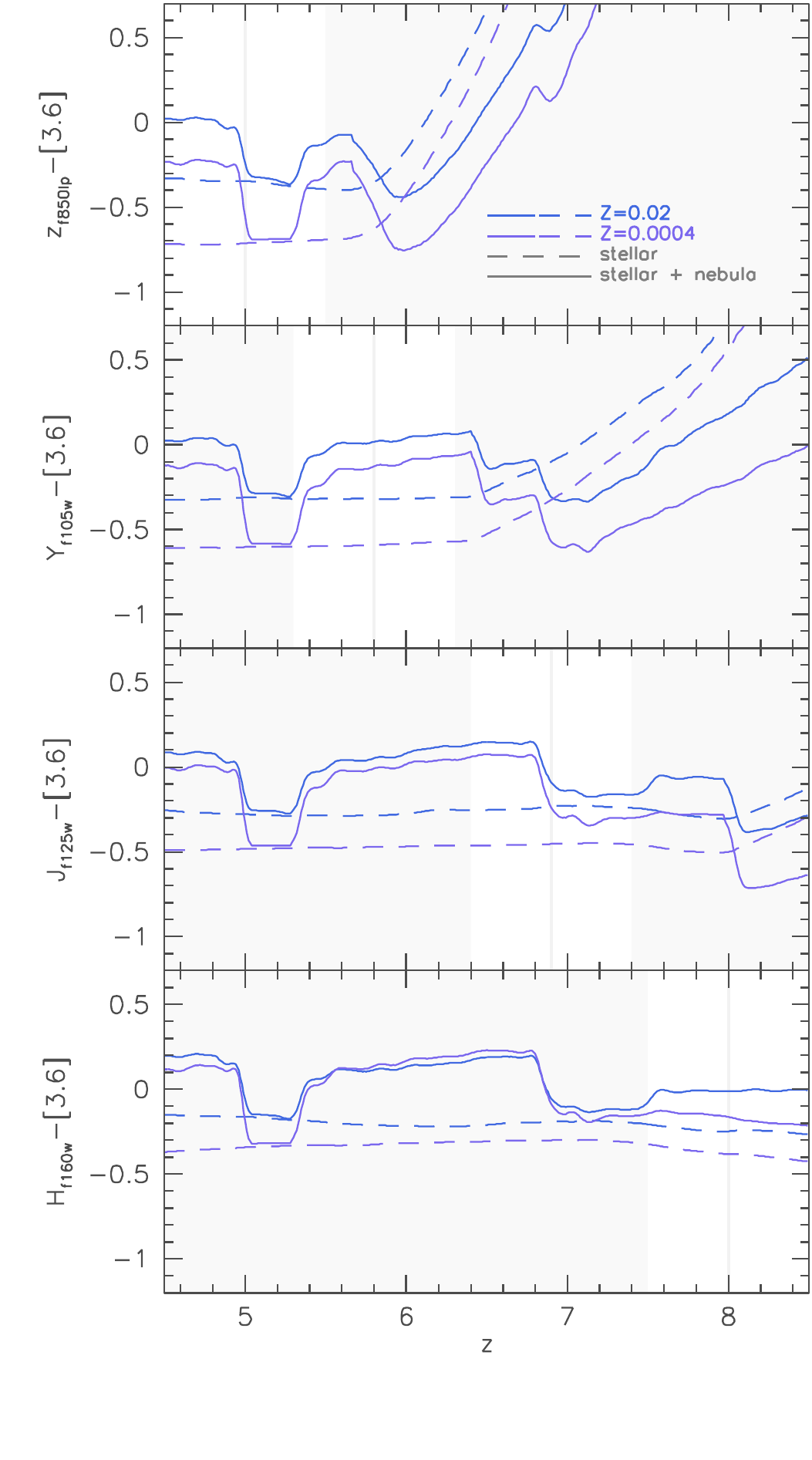}
\includegraphics[width=20pc]{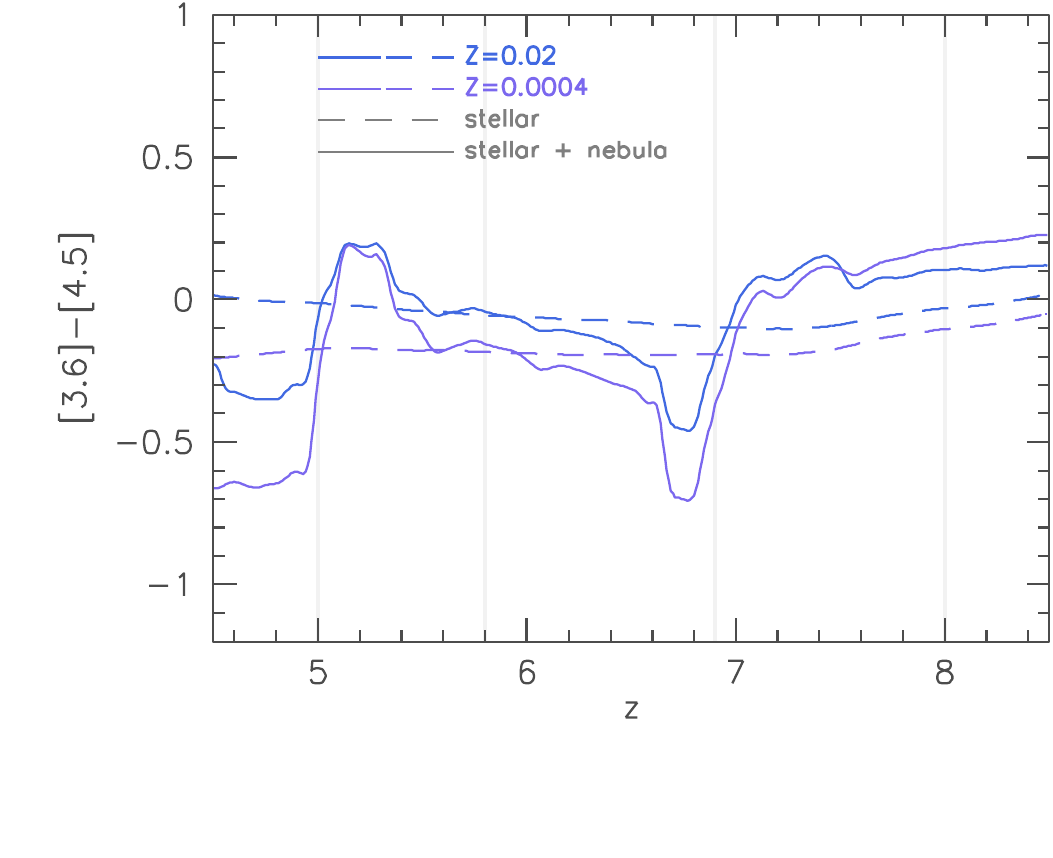}
\caption{The predicted observed-frame X$-$[3.6] (where X $\in\{z_{f850lp}$, Y$_{f105w}$, J$_{f125w}$, H$_{f160w}\}$) and [3.6]$-$[4.5] colours as a function of redshift assuming $100$ Myr previous duration of star formation for two (stellar) metallicities ($Z=0.02$ and $Z=0.0004$). The dashed lines show the result if only stellar emission is included while the solid lines show the effect of including nebular emission (continuum and line emission).}
\label{fig:Xch1z}
\end{figure}

The effect of nebular emission on observed frame colours is critically sensitive to both the filter transmission curve and the redshift of the source. Small changes in redshift can leave strong emission lines in adjacent bands, dramatically affecting the observed colour. 

This can be seen in Figure \ref{fig:Xch1z}, where the X$-$[3.6] (where X $\in\{z_{f850lp}$, Y$_{f105w}$, J$_{f125w}$, H$_{f160w}\}$) and [3.6]$-$[4.5] colours (both from the pure stellar SED and the stellar+nebular) of a young star forming stellar population ($100 {\rm Myr}$ continuous star formation) are shown as a function of redshift (Figure \ref{fig:SEDs} also shows the spectral energy distribution of a stellar population forming stars for $100\,{\rm Myr}$ along with the transmission functions of the various observed frame filters). For example, at $z=5-5.3$ there are no strong emission lines within the IRAC [3.6] bandpass, thus leaving the X$-$[3.6] colour virtually unchanged relative to that for a pure stellar SED (unless the X filter encompasses Lyman-$\alpha$). In contrast at $z<5$ and $z>5.3$ the IRAC [3.6] filter includes strong emission lines (O[III] and H$\beta$, or H$\alpha$ respectively). This results in an extremely strong sensitivity to the redshift. An increase in redshift of $0.1$ (i.e. $z=5.0\to 5.1$) can decrease the X$-$[3.6] colour by $0.5$\footnote{$z=5$ is a declining inflexion point for the X$-$[3.6] colour.} while correspondingly increasing the [3.6]$-$[4.5] colour by $0.7$. A similar situation occurs at $z\approx 7$ as the [OIII] and H$\beta$ lines move out of the [3.6] band into the [4.5] band. The [3.6]$-$[4.5] colour (which probes the rest-frame optical at $z=4-8$) also experiences significant variation as a function of redshift, as shown in the lower-panel of Figure \ref{fig:Xch1z}. Specifically, both $z=5$ and $z=6.9$ are approximately rising inflexion points, again caused by the shifting locations of the various strong emission lines.

This highlights that interpreting the observed colours of galaxies with strong emission lines is extremely challenging without precise knowledge of the redshift (or redshift distribution).


\section{Predictions from galaxy formation simulations}\label{sec:sims}

The preceding analysis demonstrated that the observed frame optical/NIR colours of high-redshift galaxies can be extremely sensitive to nebular emission. By using a galaxy formation model to predict both the star formation and metal enrichment histories we can predict the stellar spectral energy distributions (SEDs) and nebular (line and continuum) emission and thus the intrinsic observed colours.

\subsection{{\em MassiveBlack}-II}

We make use of a state-of-the-art cosmological hydrodynamic simulation of structure formation: {\em MassiveBlack}-II (for a more detailed description of this simulation see: Khandai et al. {\em in-prep}). The {\em MassiveBlack}-II simulation is performed using the cosmological TreePM-Smooth Particle Hydrodynamics (SPH) code {\sc P-Gadget}, a hybrid version of the parallel code {\sc Gadget2} (Springel 2005) tailored to run on the new generation of Petaflop scale supercomputers. {\em MassiveBlack}-II includes $N_{\rm par}=2\times 1792^{3}\approx 11.5$ billion particles in a volume of $10^{6}\,{\rm Mpc}^{3}/h^{3}$ ($100\,{\rm Mpc}/h$ on a side) and includes not only gravity and hydrodynamics but also additional physics for star formation (Springel \& Hernquist 2003), metal enrichment, black holes and associated feedback processes (Di Matteo et al.  2008, Di Matteo et al. 2012).

\subsubsection{Properties of galaxies in the simulation}

A detailed overview of the properties of galaxies (galaxy stellar mass functions, luminosity functions etc.) in the simulation is presented in Khandai et al. {\em in-prep}). Nevertheless, it is useful to present predictions for the properties which directly influence the UV-optical colours of galaxies, i.e. the star formation and metal enrichment histories.

\begin{figure}
\centering
\includegraphics[width=20pc]{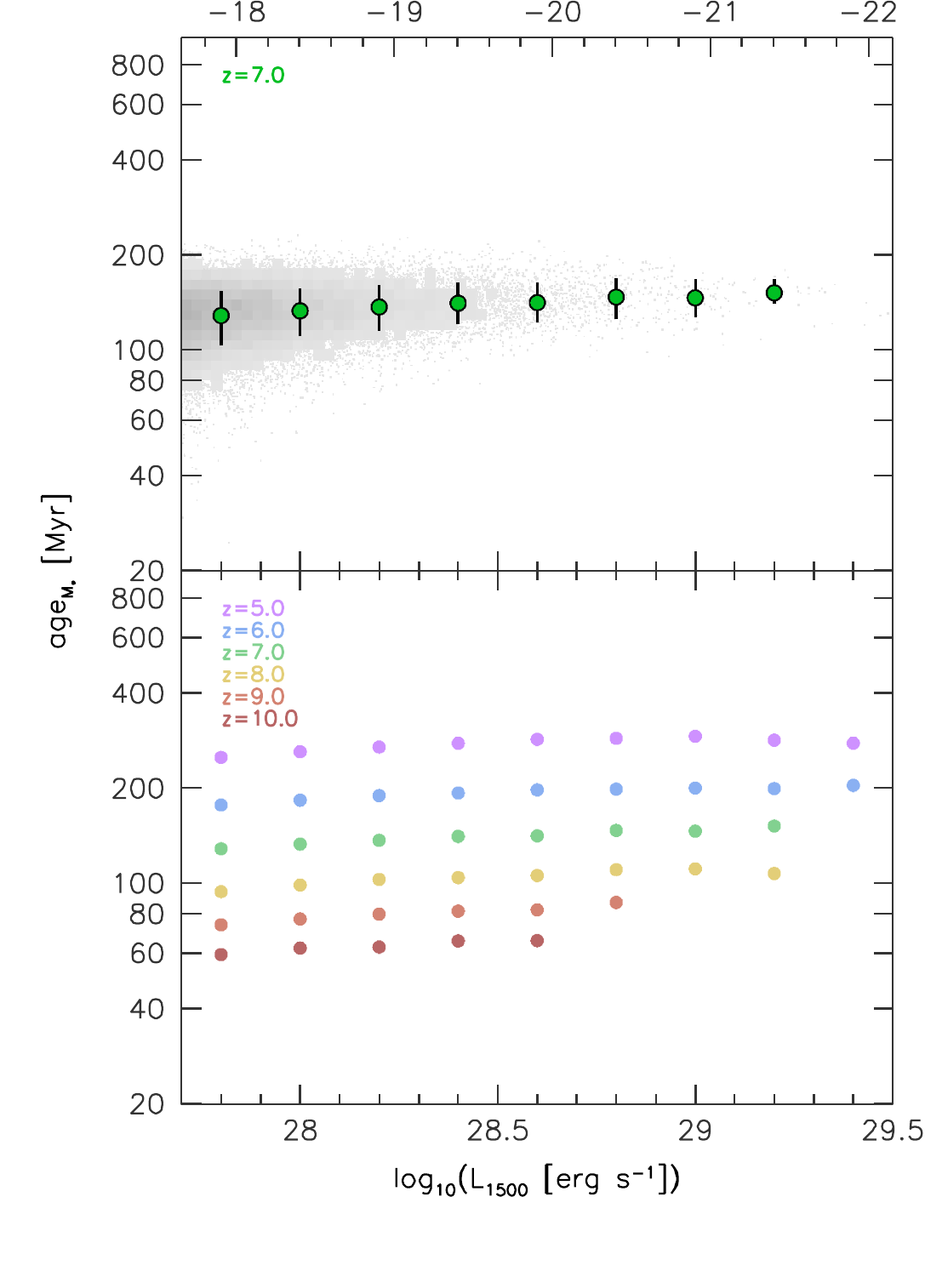}
\caption{The mass-weighted stellar age as a function of the intrinsic UV luminosity predicted by the {\em MassiveBlack}-II simulation. The upper panel shows both a density plot (with shading denoting the number of galaxies) and the median age (and $16^{\rm th}-84^{\rm th}$ percentile range) in several luminosity bins at $z=7$. The shade of the density plot denotes the number of galaxies contributing to each bin on a linear scale. Where there are fewer than 25 galaxies contributing to each bin on the density plot the galaxies are plotted individually. The lower panel shows only the median age as a function of luminosity but for a range of redshifts ($z\in\{5,6,7,8,9,10\}$).}
\label{fig:Age_lum.eps}
\end{figure}

\begin{figure}
\centering
\includegraphics[width=20pc]{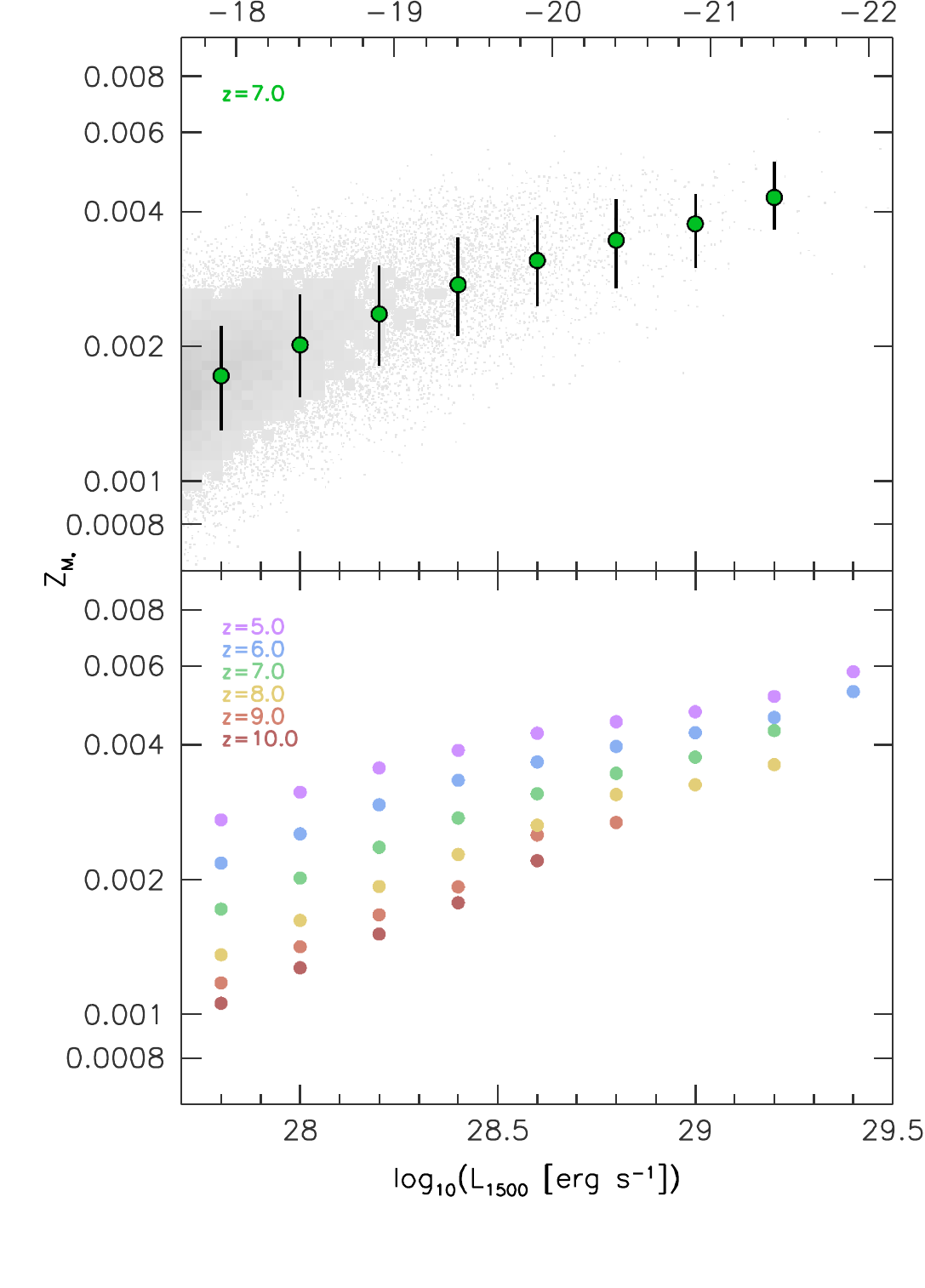}
\caption{The mass-weighted stellar metallicity as a function of the intrinsic UV luminosity predicted by the {\em MassiveBlack}-II simulation. The upper panel shows both a density plot (with shading denoting the number of galaxies) and the median metallicity (and $16^{\rm th}-84^{\rm th}$ percentile range) in several luminosity bins at $z=7$. The shade of the density plot denotes the number of galaxies contributing to each bin on a linear scale. Where there are fewer than 25 galaxies contributing to each bin on the density plot the galaxies are plotted individually. The lower panel shows only the median stellar metallicity as a function of luminosity but for a range of redshifts ($z\in\{5,6,7,8,9,10\}$).}
\label{fig:Z_lum.eps}
\end{figure}

\begin{figure}
\centering
\includegraphics[width=20pc]{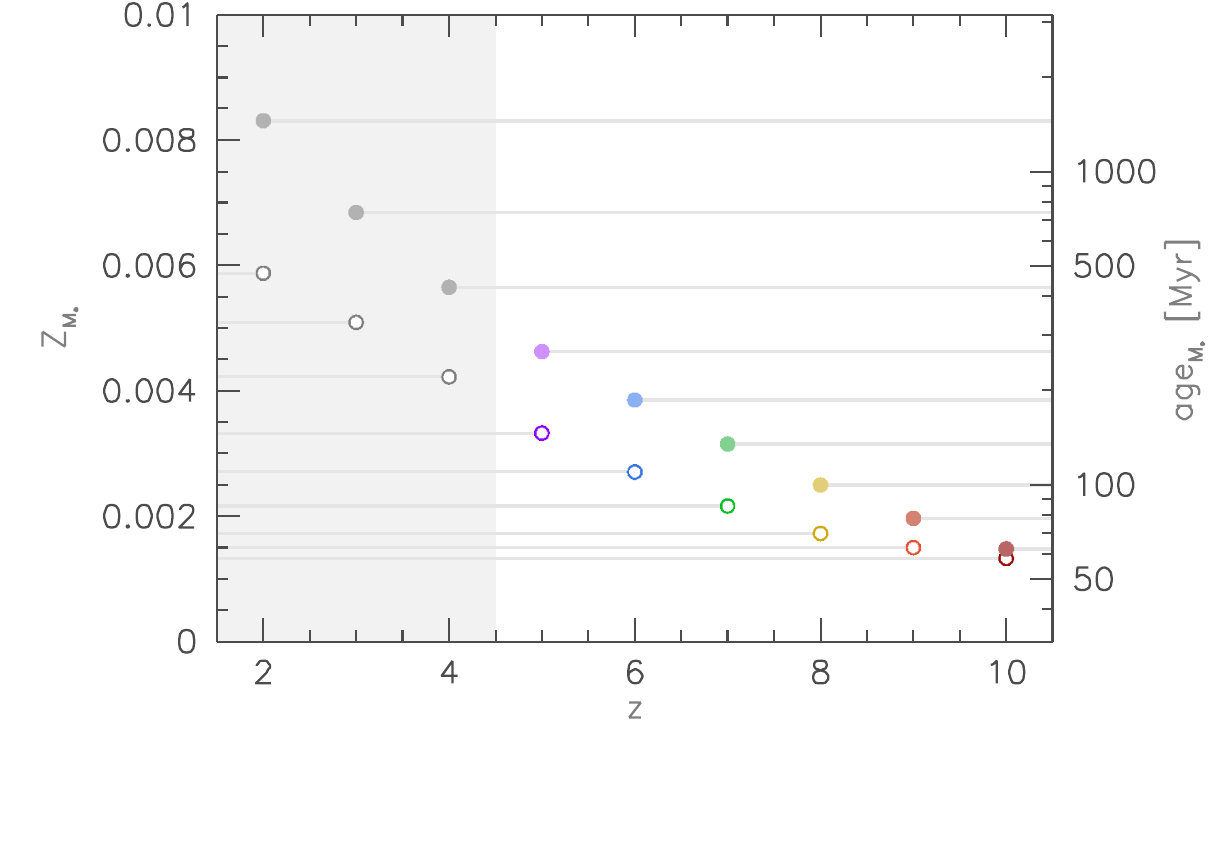}
\caption{The evolution with redshift of the simulated average (median) mass-weighted age (right-hand axis) and stellar metallicity of galaxies in the simulation with $-20.5<M_{1500}<-18.5$.}
\label{fig:ZAge_z.eps}
\end{figure}

Instead of presenting the full star formation and metal enrichment histories, in Figure \ref{fig:Age_lum.eps} we show the median mass-weighted stellar age in bins of intrinsic UV luminosity and in Figure \ref{fig:Z_lum.eps} the median mass-weighted stellar metallicity, in both cases for a range of redshifts ($z\in\{5,6,7,8,9,10\}$). The redshift trends can be seen more clearly in Figure \ref{fig:ZAge_z.eps} where the evolution of median mass-weighted stellar age and metallicity for galaxies with $-20.5<M_{1500}<-18.5$ are shown. 

Figure \ref{fig:Age_lum.eps} reveals only a very weak correlation between UV luminosity suggesting any correlation in the predicted UV-optical colours is unlikely to be dominated by variations in the star formation history. Figure \ref{fig:Z_lum.eps} on the other hand reveals a strong correlation between the mass-weighted stellar metallicity and the UV luminosity. Both quantities do however show strong variation with redshift (as can be seen clearly in Figure \ref{fig:ZAge_z.eps}).

\subsubsection{Stellar spectral energy distributions}

The stellar spectral energy distributions (SEDs) of galaxies are generated by combining the SEDs of individual star particles taking account of their metallicity and age using the stellar population synthesis (SPS) model {\sc Pegase.2} and assuming a Salpeter (1955) IMF. As noted previously, the SED predicted from a given star formation and metal enrichment history, is sensitive to the choice of SPS model. In Appendix \label{app:SPS} we consider the effect of alternative SPS models on the spectral energy distribution of galaxies predicted by the simulation.

\subsubsection{Nebular emission and dust attenuation}

Nebular emission is included in the simulated SEDs using the same prescription described in \S\ref{sec:factors.neb}. We take the ionising flux predicted from the stellar SED and determine the fluxes in the various hydrogen recombination lines assuming the escape fraction is zero. We then use the calibrations of 
Anders \& Fritze et al. (2003) to determine the fluxes in the non-hydrogen lines. We also use {\sc Pegase.2} to predict the contribution of nebular continuum emission (though this is typically very small except in extreme cases). For the results presented here we do not include dust attenuation concentrating solely on the intrinsic photometry.

\subsection{Stellar colours}

\begin{figure}
\centering
\includegraphics[width=20pc]{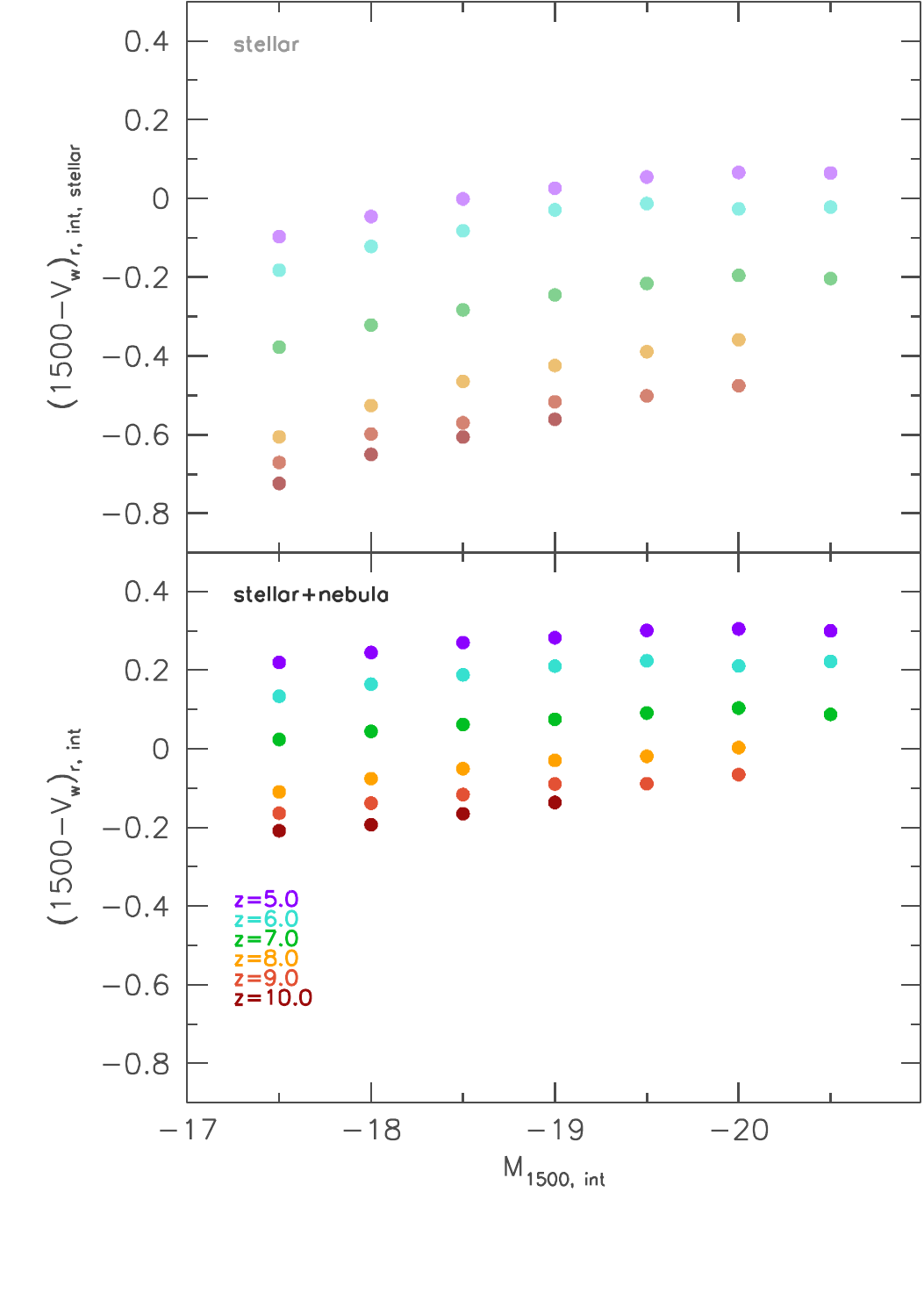}
\caption{The simulated average (median) intrinsic (i.e. with no dust attenuation) rest-frame $1500-V_{w}$ colours of galaxies as a function of luminosity for galaxies at $z\in\{5,6,7,8,9,10\}$. The upper-panel shows only the pure stellar colours while the lower panel shows the average colour including the effects of nebular emission.}
\label{fig:MB_1500V}
\end{figure}

\begin{figure}
\centering
\includegraphics[width=20pc]{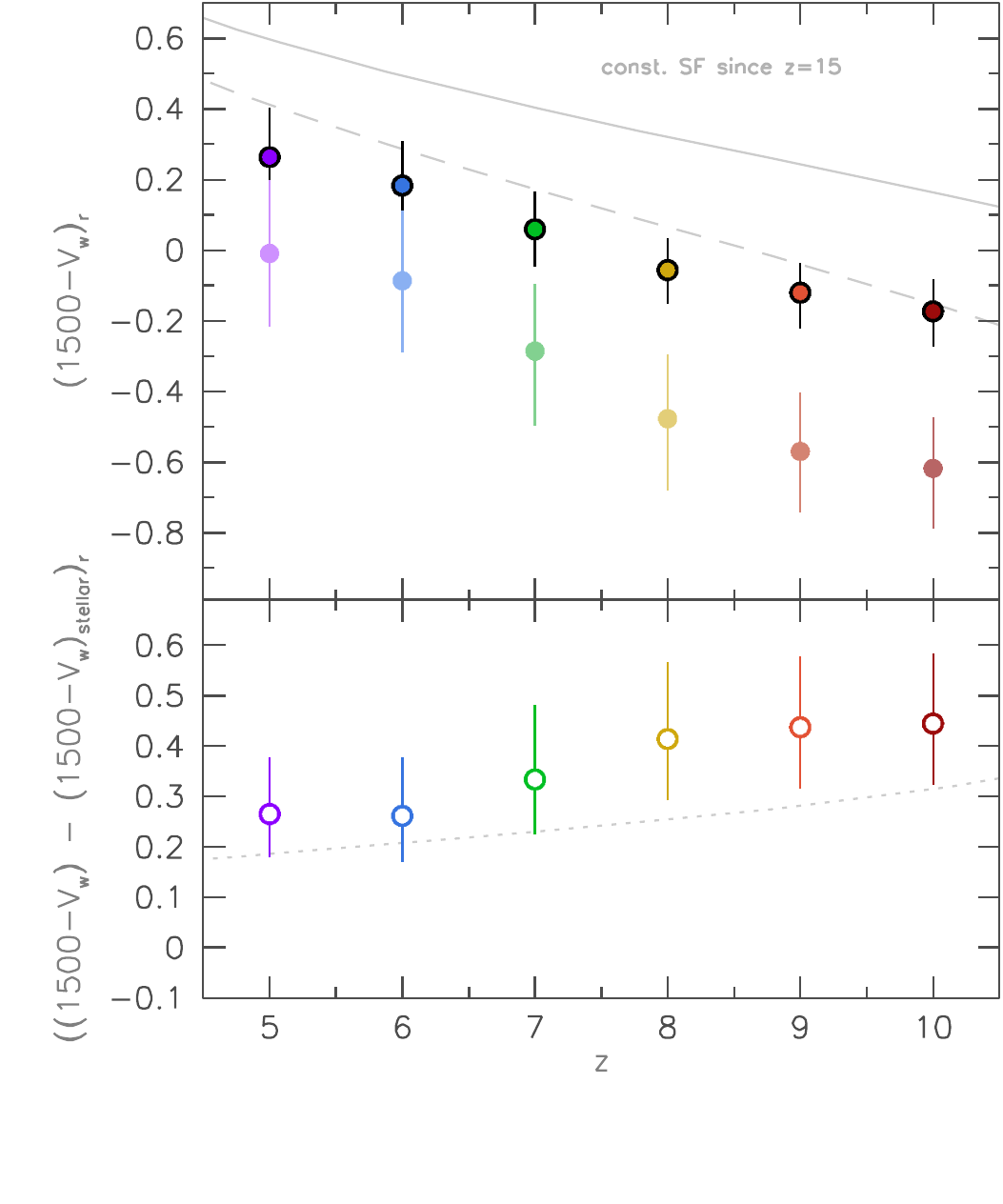}
\caption{The redshift evolution of the median rest-frame $1500-V_{w}$ colour (points) and $16^{\rm th}-84^{\rm th}$ percentile range for both the simulated pure stellar SED and the simulated SED including nebular emission (outlined points) for galaxies with $-20.5<M_{1500}<-18.5$.  The lower panel shows the difference between the pure stellar colour and the colour including nebular emission. The dashed and solid lines show the prediction (using the {\sc Pegase.2} SPS model) assuming constant star formation since $z=15$ for the pure stellar and stellar with nebular emission SEDs respectively.}
\label{fig:MB_1500V_z}
\end{figure}

The top panel of Figure \ref{fig:MB_1500V} shows the median rest-frame pure stellar $1500-V_{w}$ colour as a function of the intrinsic UV luminosity ($M_{1500,{\rm int}}$) for galaxies at $z\in\{5,6,7,8,9,10\}$. The $1500-V_{w}$ colour is correlated, albeit weakly, with the UV luminosity with the colour reddening by $\sim 0.15$ as $M_{1500,{\rm int}}=-18\to -20$ (at $z=7$). The $1500-V_{w}$ colour is also strongly correlated with redshift as can be seen more clearly in Figure \ref{fig:MB_1500V_z} where the median pure stellar colour (of galaxies with $-20.5<M_{1500}<-18.5$) is shown as a function of redshift. For example, the median $1500-V_{w}$ colour increases by $\sim 0.5$ from $z=8\to 5$. 

The distribution of pure stellar colours for galaxies with $-20.5<M_{1500}<-18.5$ is shown in Figure \ref{fig:distr_colours}. The $16^{\rm th}-84^{\rm th}$ percentile range is $~0.15\,{\rm mag}$ though this increases slightly to lower redshift.

The correlation with redshift is driven by the variation in both the average star formation and metal enrichment histories of galaxies (as shown in Figure \ref{fig:ZAge_z.eps}) while the correlation with luminosity is driven predominantly by the variation in the metal enrichment history.

\subsection{The effect of nebular emission}\label{sec:sims.neb}

As noted in \S\ref{sec:factors.neb} the effect of nebular emission will, be adding additional flux in the $V_{w}$-band, have the result of reddening the rest-frame $1500-V_{w}$ colour relative to the pure stellar colour. This is shown for our simulated galaxies in Figures \ref{fig:MB_1500V} and \ref{fig:MB_1500V_z}. 

As can be seen in Figures \ref{fig:MB_1500V} and \ref{fig:MB_1500V_z} the effect of adding nebular emission is to flatten the correlation between the $1500-V_{w}$ colour and the UV luminosity and redshift. The median $1500-V_{w}$ colour including nebular emission only reddens by $\sim 0.05$ from $M_{1500,{\rm int}}=-18\to -20$ (at $z=7$). This is because the relative strength of nebular emission is inversely correlated with both stellar metallicity and age. Those galaxies with bluer stellar colours, which are indicative of more recent star formation or lower metallicity (which are more common at higher redshift and lower luminosity) will then have stronger nebular emission and consequently will be reddened by nebular emission more than those with redder stellar colours. This has the effect of diminishing the correlation of the $1500-V_{w}$ colour with redshift and luminosity. The median $1500-V_{w}$ colour including nebular emission only reddens by $\sim 0.1$ from $M_{1500,{\rm int}}=-18\to -20$ and $\sim 0.3$ from $z=8\to 5$ (c.f. $0.1$ and $0.3$ for the pure stellar case respectively). 

\subsubsection{The distribution of colours}

\begin{figure}
\centering
\includegraphics[width=20pc]{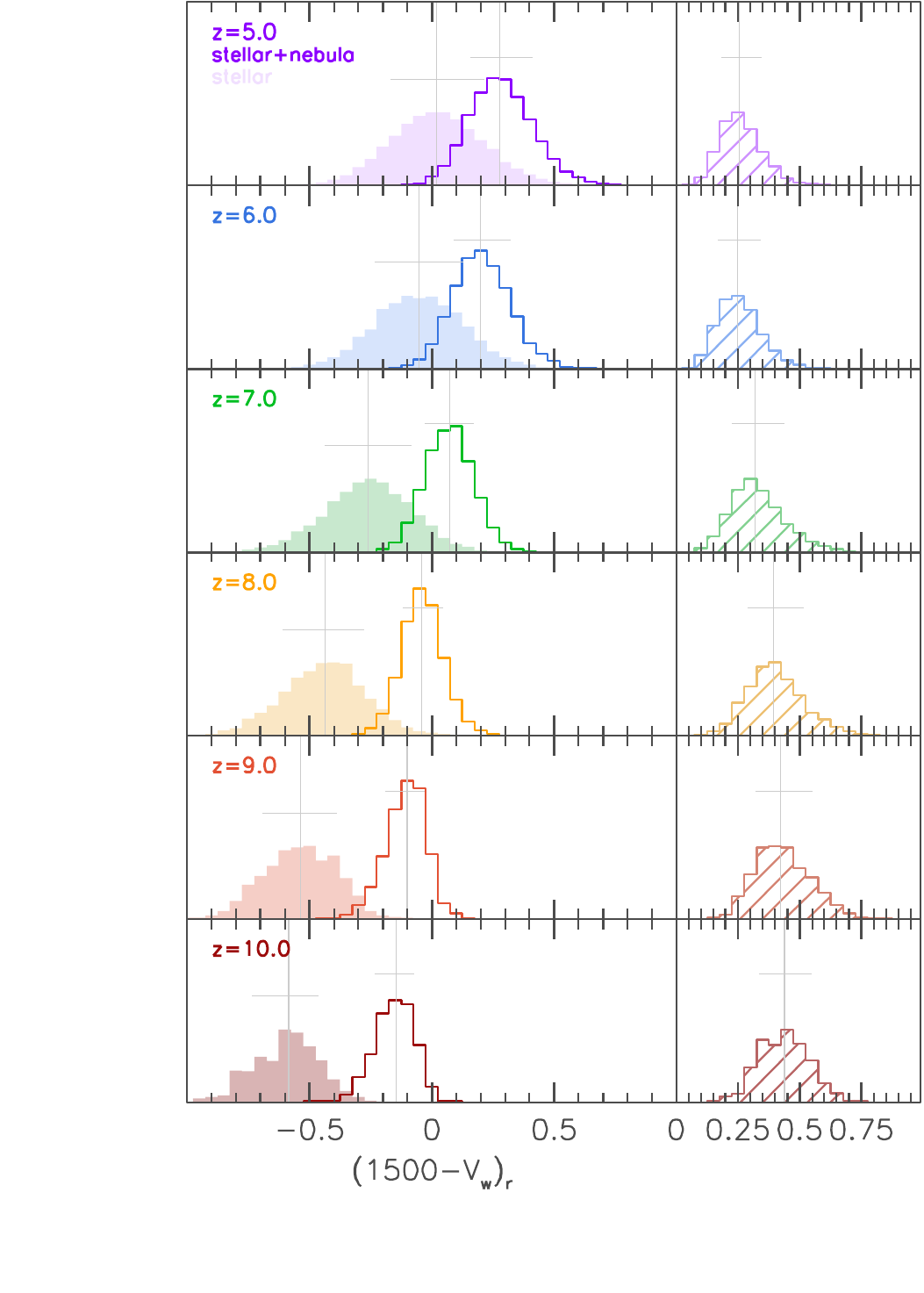}
\caption{{\em left-hand panels} - The distribution of both pure stellar (shaded histogram) and stellar with nebular (line histogram) intrinsic rest-frame $1500-V_{w}$ colours for galaxies with $-20.5<M_{1500}<-18.5$ at $z\in\{5,6,7,8,9,10\}$. {\em right-hand panels} - The distribution of colour residuals. In each case the vertical and horizontal lines denote the median and $16^{\rm th}-84^{\rm th}$ percentile range of the distribution respectively.}
\label{fig:distr_colours}
\end{figure}

Because galaxies with bluer pure stellar galaxies typically have lower metallicities and/or ages (and consequently stronger nebular emission) the inclusion of nebular emission reduces the scatter in the $1500-V_{w}$ colour (as measured by the $16^{\rm th}-84^{\rm th}$ percentile range), as can be seen in Figure \ref{fig:distr_colours}. At $z=8$ this reduces the scatter (as measured by the $16^{\rm th}-84^{\rm th}$ percentile range) by almost a factor of $\times 2$ while at lower redshift it is less important.

\subsubsection{The distribution equivalent widths}

\begin{figure}
\centering
\includegraphics[width=20pc]{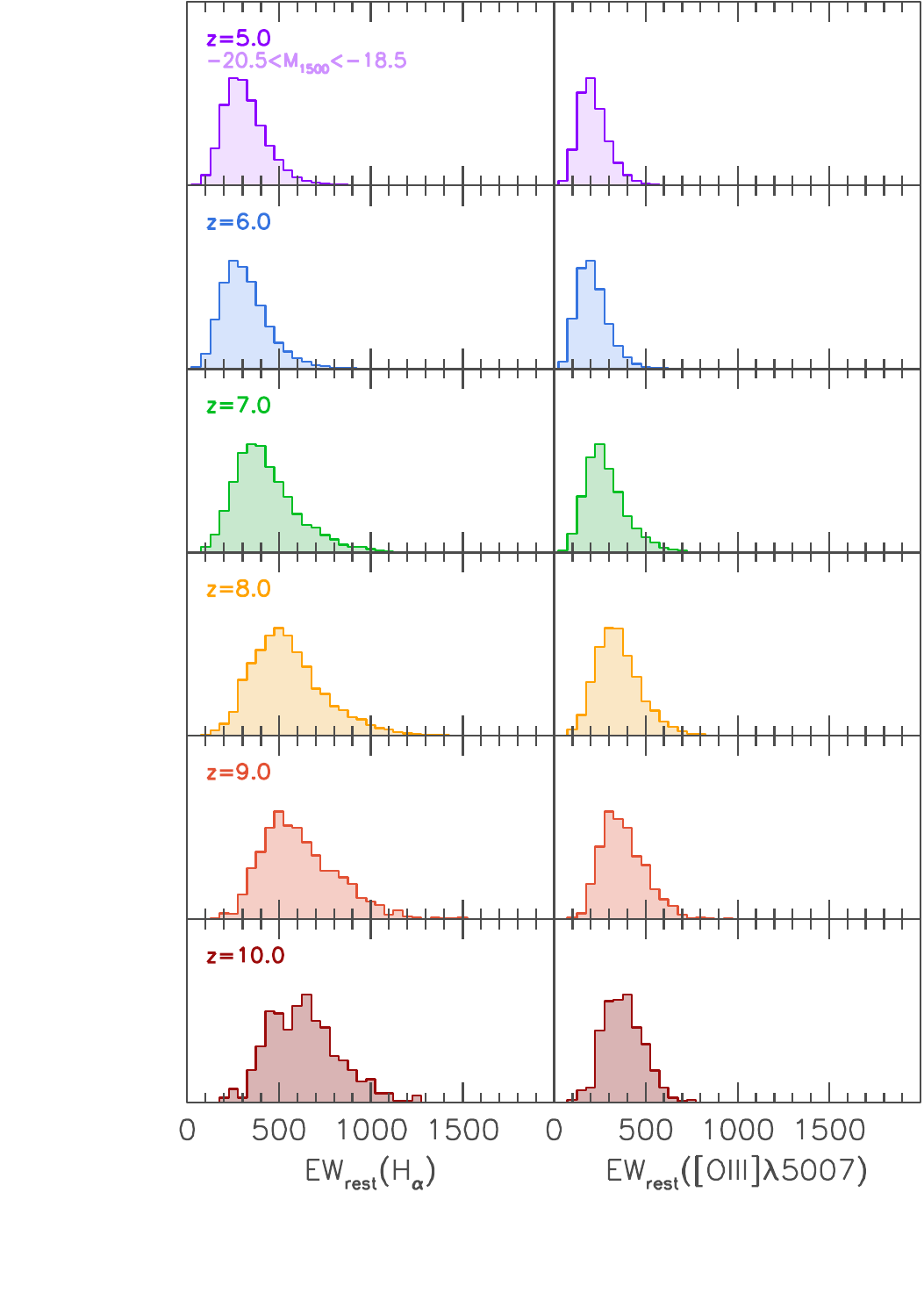}
\caption{The distribution of rest-frame intrinsic equivalent widths of the H$\alpha$ (left) and [OIII]$\lambda 5007$ predicted by the simulation for galaxies with $-20.5<M_{1500}<-18.5$ at $z\in\{5,6,7,8,9,10\}$.}
\label{fig:EW_dist}
\end{figure}

One alternative measure of the relative strength of nebular emission is the distribution of the equivalent widths of the prominent H$\alpha$ and [OIII]$\lambda 5007$ emission lines. These are shown in Figure \ref{fig:EW_dist} and follow a similar patten to the distribution of colour increments shown in Figure \ref{fig:distr_colours}.

\subsubsection{The effect on observed colours}

\begin{figure}
\centering
\includegraphics[width=20pc]{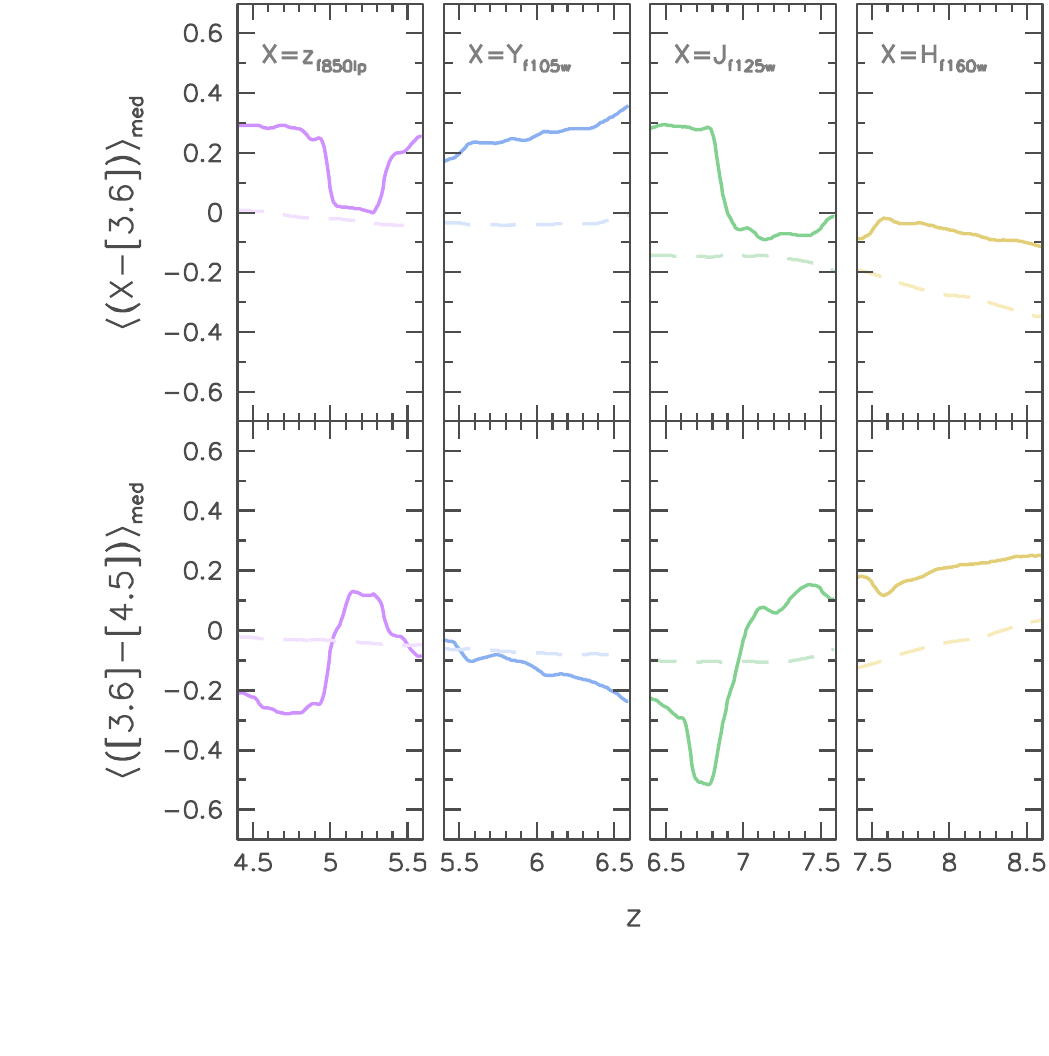}
\caption{The simulated observed frame X$-$[3.6] ({\em top panels}, X $\in\{z_{f850lp}$, Y$_{f105w}$, J$_{f125w}$, H$_{f160w}\}$) and [3.6]$-$[4.5] colours as a function of redshift across a narrow range centred on the snapshot redshift. The dashed and solid lines again denote the colours for the pure stellar SEDs and those including nebular emission respectively.}
\label{fig:MB_ObsC}
\end{figure}

As noted in \S\ref{sec:factors.neb}, the effect of nebular emission on observed-frame colours is strongly sensitive to both the choice of filters and the redshift of the source. Even a small change in redshift can have a dramatic effect on the observed colour. Figure \ref{fig:MB_ObsC} shows the average simulated (for galaxies with $-20.5<M_{1500}<-18.5$) observed frame X$-$[3.6] and [3.6]$-$[4.5] colours for redshift ranges centred on the median redshift of observed drop-out samples (see Labbe et al. 2010, Gonzalez et al. 2011, Labbe et al. 2012): V$_{f606w}$-band drop out: $\langle z\rangle= 5.0$, $i_{f775w}$: $\langle z\rangle= 5.9$, $z_{f850lp}$: $\langle z\rangle= 6.9$, and Y$_{f105w}$: $\langle z\rangle= 8.0$. In each case the X filter is chosen (from the available {\em Hubble} ACS and WFC3 filters) such that it {\em approximately} probes the rest-frame UV continuum at $1500{\rm\AA}$. Therefore we use: $z\approx5\to$ $z_{f850lp}$, $z\approx 6\to$ Y$_{f105w}$, $z\approx 7\to$ J$_{f125w}$, and $z\approx 8\to$ H$_{f160w}$.

The trends seen in Figure \ref{fig:MB_ObsC} closely reflect those demonstrated in \S\ref{sec:factors.neb} (and seen in Figure \ref{fig:Xch1z}). We see at $z\approx 5$ and $z\approx 7$ the observed colours are particularly sensitive to the redshift, with changes in redshift of $0.1$ changing the {\em average} observed colours by up to $0.3\,{\rm mag}$.

Stark et al. (2012, hereafter S12) studied the observed frame $[3.6]-[4.5]$ colours of a spectroscopically confirmed sample of galaxies at $3<z<7$. S12 compare the $[3.6]-[4.5]$ colour distribution at $3.8<z<5.0$ (where H$\alpha$ emission, if present, will contaminate the $[3.6]$ filter) with those at $3.1<z<3.6$ (where there are expected to be no strong nebular emission lines in either the $[3.6]$ or $[4.5]$ filters) finding the $[3.6]-[4.5]$ colour at $3.8<z<5.0$ is $0.33\,{\rm mag}$ bluer than at $3.1<z<3.6$. This is interpreted as being due to the presence of nebular line emission. This closely matches our prediction at $4.5<z<4.9$ where we find the effect of nebular emission results in the observed $[3.6]-[4.5]$ colour being $0.25\,{\rm mag}$ bluer than the pure stellar colour. S12 also attempt to determine the increase in the $[3.6]$ flux due to nebular emission at $3.8<z<5.0$ by comparing the observed $[3.6]$ flux with that expected from the best fit stellar continuum models, and find an increase in the $[3.6]$ flux ($[3.6]_{\rm neb}-[3.6]_{\rm stellar}=-0.27\,{\rm mag}$) consistent with our predictions at $4.5<z<4.9$ ($[3.6]_{\rm neb}-[3.6]_{\rm stellar}=-0.29\,{\rm mag}$).

\section{The effect on stellar mass estimates}\label{sec:stellar_masses}

We have seen that the effect of nebular emission is to typically redden the observed colour probing the rest-frame UV-optical relative to the pure stellar colour. For individual galaxies at $z\approx 7$ this can be as large as $+0.6\,{\rm mag}$ with the average at $z=6.8$ being $+0.4\,{\rm mag}$ for J$_{f125w}-$[3.6]. The effect is also strongly dependent on the redshift and choice of observed filters with changes in the redshift of as little as $\pm 0.1$ is able to change the observed colours by $>0.3\,{\rm mag}$. 

Because accurate estimates of the stellar mass-to-light ratio require a measurement of the shape rest-frame UV-optical SED (e.g. Wilkins et al. 2013b) the effect of nebular emission can introduce a large bias, particularly where the redshift is not known accurately. The exact size of the effect of nebular emission on stellar mass estimates however depends on various factors including the method used to measure the mass-to-light ratio, the redshift. 

\begin{figure}
\centering
\includegraphics[width=20pc]{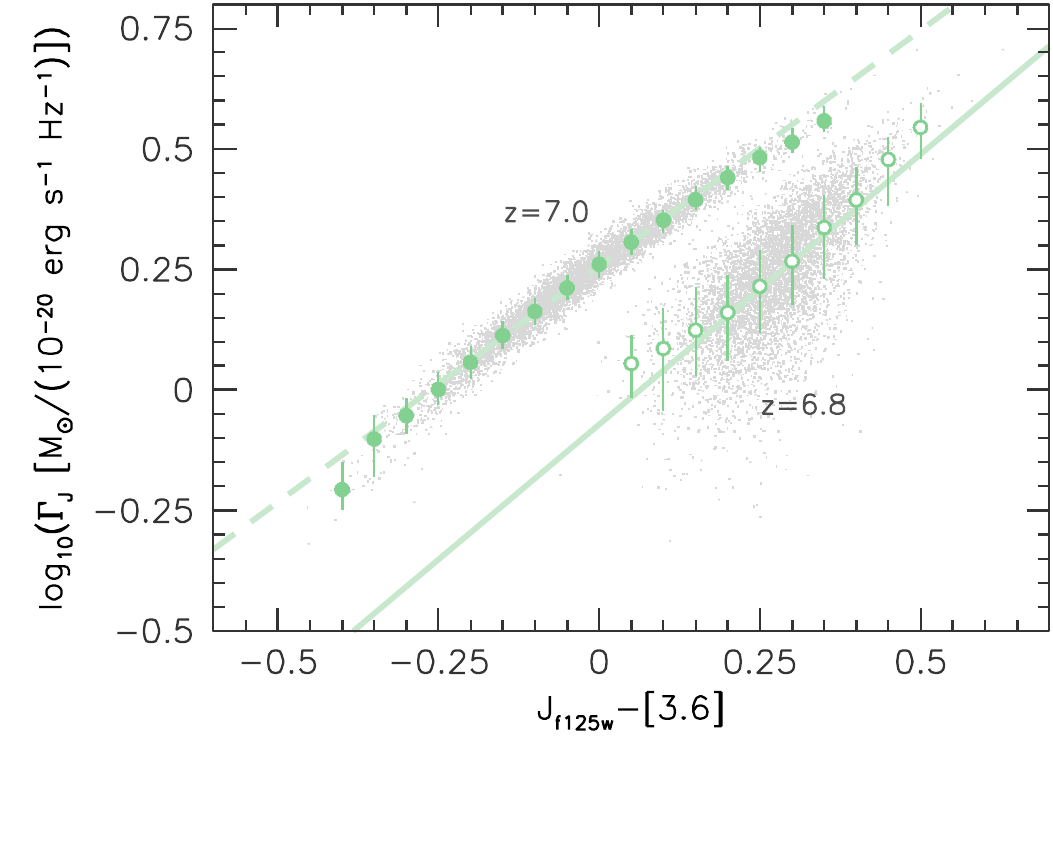}
\caption{The simulated observed frame J$_{f125w}-$[3.6] colours and J$_{f125w}$-band mass-to-light ratios of galaxies at $z=6.8$ and $z=7.0$ with $-20.5<M_{1500}<-18.5$. The two straight lines denote simple linear fits to the two sets of simulations.}
\label{fig:MTOL}
\end{figure}

As a simple illustration we consider the case where the stellar mass-to-light ratio is measured using a single colour (e.g. Taylor et al. 2012, Wilkins et al. 2013). Figure \ref{fig:MTOL} shows the relationship between the observed frame J$_{f125w}-$[3.6] colours and  J$_{f125w}$-band stellar mass-to-light ratios of galaxies at $z=6.8$ and $z=7.0$. In both cases the J$_{f125w}-$[3.6] colour is correlated with the mass-to-light ratio (though the correlation is much stronger at $z=7.0$ where the effect of nebular emission is smaller than at $z=6.8$), however, for the same observed colour the average mass-to-light ratio at $z=7.0$ is $\sim 0.25\,{\rm dex}$ ($1.8\times$) larger. That is, were the redshift erroneously assumed to be at $z=7.0$ instead of the true redshift of $z=6.8$ the stellar mass-to-light ratio would be overestimated by around $\sim 0.25\,{\rm dex}$.

\section{Conclusions}\label{sec:c}

We have explored the evolution of the rest-frame UV/optical (and observed frame near-IR) colours of high-redshift galaxies ($z=5-10$) predicted by our large cosmological hydrodynamical simulation {\em MassiveBlack}-II. 

We find that the median rest-frame {\em pure} stellar UV-optical ($1500-V_{w}$) colour is correlated with both luminosity and redshift. The $1500-V_{w}$ colour reddens by $\sim 0.2$ as the luminosity increases from $M_{1500}=-18$ to $-20$ (at $z=7$) and by $\sim 0.5$ from $z=8$ to $z=5$. In both cases, this reflects the trend of increasing ages and metallicities with luminosity and to lower redshift.  

However, when nebular emission is included, these correlations weaken. The $1500-V_{w}$ colour reddens by only $\sim 0.3\,{\rm mag}$ as $z=8\to 5$ and $\sim 0.1\,{\rm mag}$ as $M_{1500}=-18\to -20$. This occurs because galaxies with very blue stellar colours (indicative of galaxies with recent star formation or low metallicity) typically have stronger nebular emission causing their colours to redden by a greater relative amount. 

The effect of nebular emission on observed frame colours is very sensitive to both the choice of filters and redshift. For example, at $z=7.0$, nebular emission only reddens the observed J$_{f125w}-$[3.6] colour by, on average, $\sim 0.1\,{\rm mag}$ while at $z=6.8$ it reddens the J$_{f125w}-$[3.6] colour by $\sim 0.45\,{\rm mag}$ (i.e. a difference of $\sim 0.35\,{\rm mag}$). Similarly, at $z=7.1$, nebular line emission causes the [3.6]$-$[4.5] to redden by $\sim 0.2\,{\rm mag}$, while at $z=6.8$ it causes [3.6]$-$[4.5] colour to shift blue-ward by $\sim 0.4\,{\rm mag}$ (a net difference of $\sim 0.6\,{\rm mag}$). 

This strong sensitivity of observed colours to the redshift makes interpreting the colours of individual objects extremely difficult unless precise redshifts are known. Indeed, if the stellar mass-to-light ratio were inferred from the J$_{f125w}-$[3.6] colour alone a difference of $\pm 0.35\,{\rm mag}$ (as expected between $z=6.8$ and $7.0$) would roughly double 

While the general trends we observe hold true irrespective of the choice of initial mass function and stellar population synthesis model both these factors can strongly affect the predicted colours. For example, utilising the Maraston et al. (2005) model yields stellar colours between $0.1-0.3\,{\rm mag}$ redder than the {\sc Pegase.2} model (which is assumed throughout this work).

\subsection*{Acknowledgements}

We would like to thank the anonymous referee for their useful comments and suggestions that we feel have greatly improved this manuscript. SMW and AB acknowledge support from the Science and Technology Facilities Council. WRC acknowledges support from an Institute of Physics/Nuffield Foundation funded summer internship at the University of Oxford. RACC thanks the Leverhulme
Trust for their award of a Visiting Professorship at the University of Oxford. The simulations were run on the Cray XT5 supercomputer Kraken at the National Institute for Computational Sciences. This research has been funded by the National Science Foundation (NSF) PetaApps program, OCI-0749212 and by NSF AST-1009781.

\bsp

\appendix\label{app:SPS}
\section{The choice of Stellar Population Synthesis model}

A key ingredient in our analysis is the transformation of the simulated star formation and metal enrichment history into a spectral energy distribution through the use of stellar population synthesis (SPS) modelling. Stellar population synthesis models work by combining stellar tracks, which model different stellar evolution phases, with spectral libraries, which empirically or theoretically relate the spectral output to individual stars based on their mass, age, and chemical composition. By combining these with a choice of initial mass function and initial chemical composition it then becomes possible to model the spectral energy distributions (SEDs) of simple stellar populations.

\begin{figure}
\centering
\includegraphics[width=20pc]{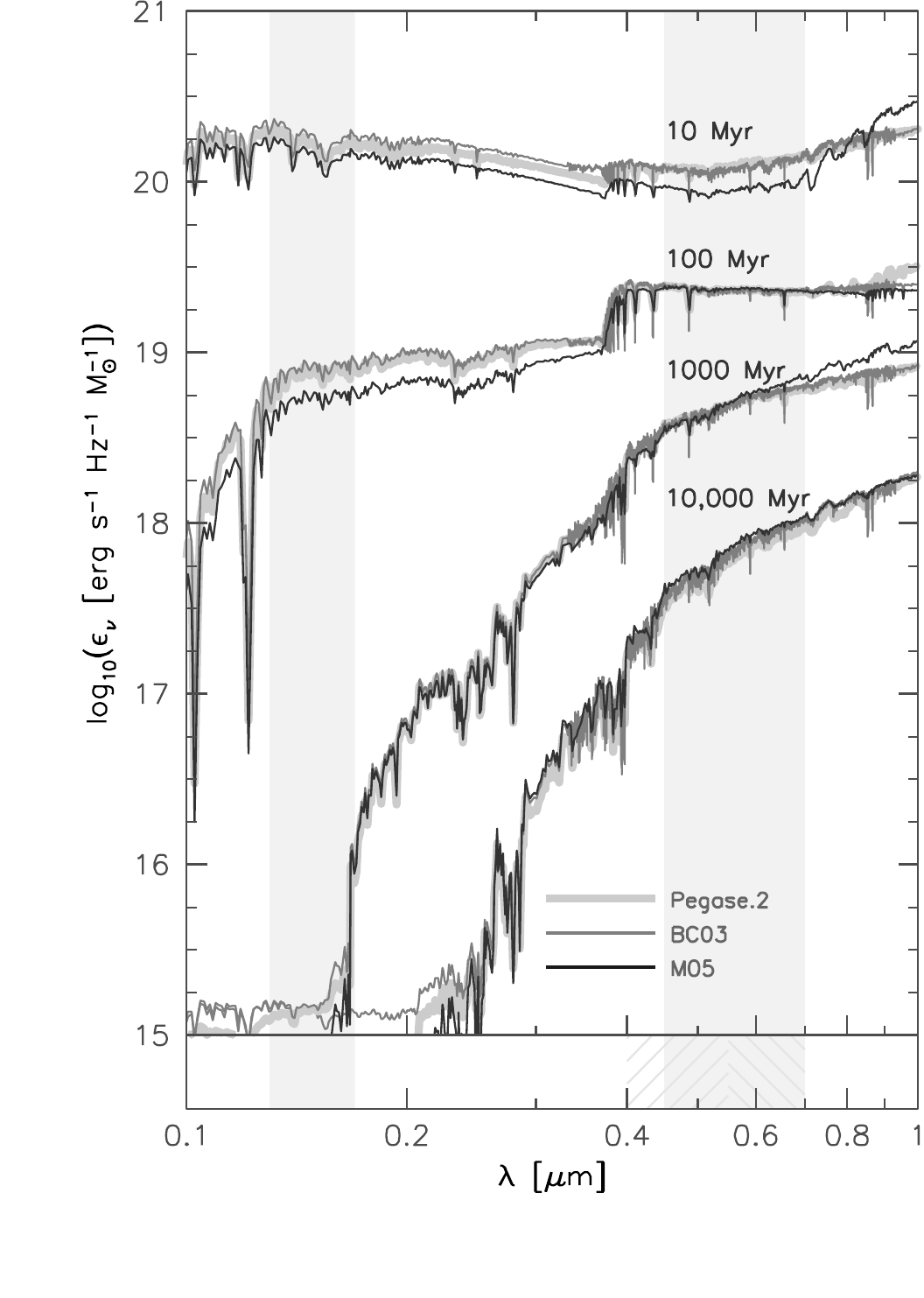}
\caption{The pure stellar spectral energy distribution of simple stellar populations (with age $\in\{10,100,1000,10000\}$ Myr and $Z=0.02$) predicted by various SPS models. Each SED is normalised to have the same initial mass. The two vertical shaded bands denote the location and width of the $1500$ and $V_{w}$ filters considered in this work.}
\label{fig:SPS_SED}
\end{figure}

However, due to differences in the treatment of stellar evolution and the utilised spectral libraries, SPS models produce varying results for the spectral energy distributions of stellar populations (see, for example, Conroy \& Gunn 2010). An example of some of these differences can be seen in Figure \ref{fig:SPS_SED} where we show the SEDs of simple stellar populations with ages between $10\,{\rm Myr}$ and $10\,{\rm Gyr}$ predicted assuming three popular SPS models: {\sc Pegase.2}; BC03: Bruzual \& Charlot (2003); M05: Maraston et al. (2005). For the youngest ages considered ($10$ and $100\,{\rm Myr}$) there are significant differences between the models.

For older populations ($>1\,{\rm Gyr}$) the models are relatively consistent over the rest-frame UV-optical. In the near-IR however the M05 model predicts an excess of flux at $1\,{\rm Gyr}$, attributed to a more detailed treatment of the TP-AGB stage. While this produces significant enhancement of the near-IR flux it will have little effect on this work due to our focus on rest-frame UV-optical colours and very-high redshift galaxies.

\begin{figure}
\centering
\includegraphics[width=20pc]{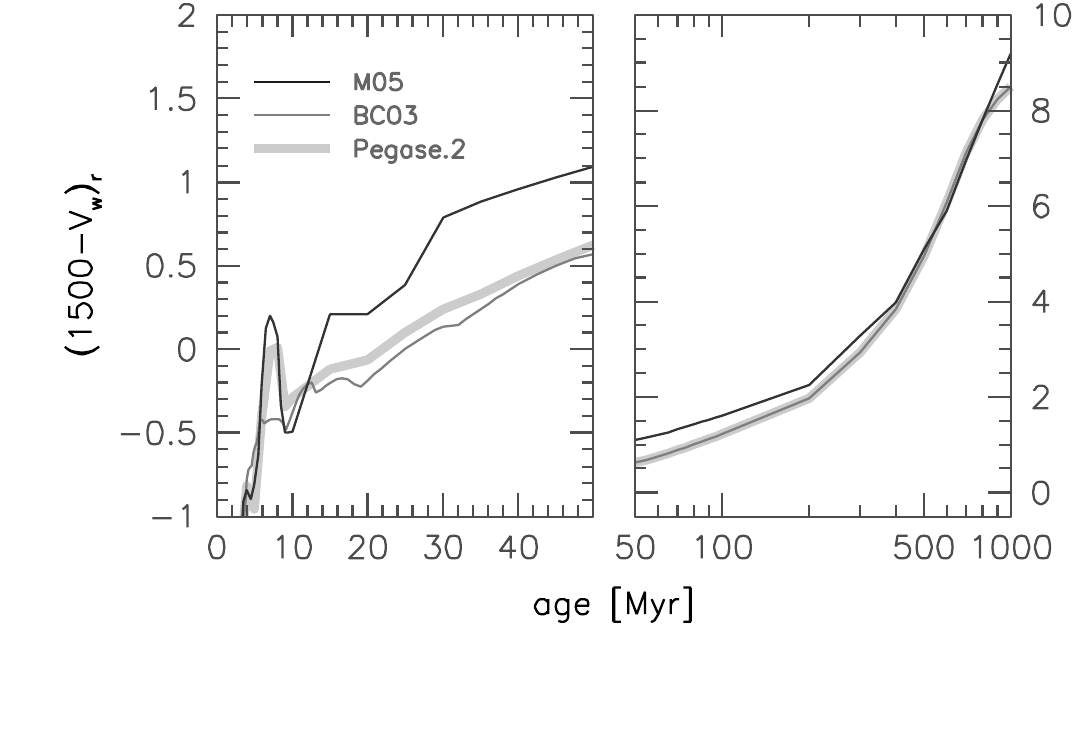}
\caption{The evolution of the rest-frame stellar $1500-V_{w}$ as a function of age for a simple stellar population ($Z=0.02$) assuming the M05, BC03, and {\sc Pegase.2} SPS models.}
\label{fig:SPS_1500V}
\end{figure}

The variation between different SPS models, and its relevance to this work, can be seen more clearly in Figure \ref{fig:SPS_1500V} where we show the evolution of the rest-frame stellar $1500-V_{w}$ as a function of age for a simple stellar population (with $Z=0.02$). While the BC03 and {\sc Pegase.2} models yield a similar colour evolution the M05 model predicts significantly redder colours at ages $10-500\,{\rm Myr}$.

\begin{figure}
\centering
\includegraphics[width=20pc]{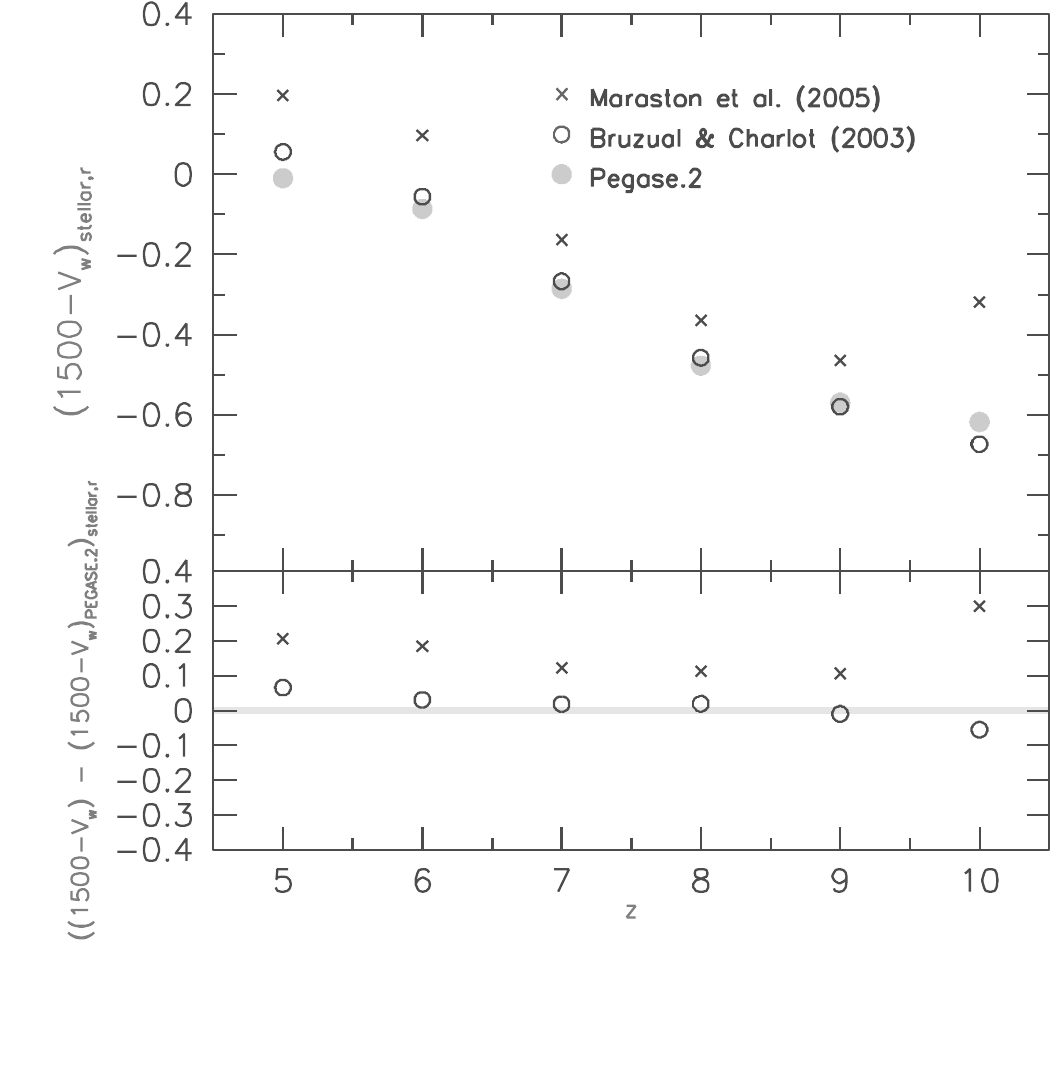}
\caption{The median pure stellar intrinsic rest-frame $1500-V_{w}$ colour of galaxies with $-20.5<M_{1500}<-18.5$ at $z\in\{5,6,7,8,9,10\}$ determined assuming various SPS models (c.f. Figure \ref{fig:MB_1500V_z}). The lower panel shows the difference between the default {\sc Pegase.2} predictions and the M05 and BC03 model predictions.}
\label{fig:SPS_1500Vw_z}
\end{figure}

This variation between different models will then leaves the colours predicted by our analysis of the {\em MassiveBlack}-II simulation sensitive to the choice of model. In Figure \ref{fig:SPS_1500Vw_z} we show the predicted rest-frame stellar $1500-V_{w}$ colour assuming various SPS models (c.f. Figure \ref{fig:MB_1500V_z}); the lower panel of this figure shows the difference between the alternative models considered and the {\sc Pegase.2} model utilised throughout this work. While the use of the BC03 model yields colours similar ($<|0.1|\,{\rm mag}$ difference) to those assuming the {\sc Pegase.2} (default) model the M05 model yields colours which are typically between $0.1-0.2\,{\rm mag}$ redder. At the most extreme, use of the M05 model at $z=10$ yields colours $\approx 0.3\,{\rm mag}$ redder than using {\sc Pegase.2}.

\end{document}